\documentclass[twocolumn]{emulateapj}

\shorttitle{X-ray Observations of Abell 2626}
\shortauthors{Wong et al.}
\slugcomment{Astrophysical Journal, accepted}

\begin{document}

\title{XMM-Newton and Chandra Observations of Abell 2626: \\
Interacting Radio Jets and Cooling Core with Jet Precession?}

\author{Ka-Wah Wong\altaffilmark{1},
Craig L. Sarazin\altaffilmark{1},
Elizabeth L. Blanton\altaffilmark{2},
and
Thomas H. Reiprich\altaffilmark{3}}

\altaffiltext{1}{Department of Astronomy, University of Virginia, 
P.~O.~Box 400325, Charlottesville, VA 22904-4325, USA; 
kwwong@virginia.edu, 
sarazin@virginia.edu}
\altaffiltext{2}{Institute for Astrophysical Research, Boston University, 
725 Commonwealth Ave., Boston, MA 02215, USA; eblanton@bu.edu}
\altaffiltext{3}{Argelander-Institut f\"ur Astronomie, Universit\"at 
Bonn, Auf dem H\"ugel 71, D-53121 Bonn, Germany; thomas@reiprich.net}

\begin{abstract}
We present a detailed analysis of the {\it XMM-Newton} and {\it Chandra} 
observations of Abell~2626 focused on the X-ray and radio interactions.
Within the region of the radio mini-halo ($\sim 70$~kpc), there
are substructures which are probably produced by the central 
radio source and 
the cooling core.
We find that there is no obvious correlation between the radio bars and 
the X-ray image.
The morphology of Abell~2626 is more complex than that of 
the standard X-ray radio bubbles seen in other cool core clusters.
Thus, Abell~2626 provides a challenge to models for the
cooling flow -- radio source interaction.
We identified two soft X-ray (0.3--2~keV) peaks with the two central 
cD nuclei;
one of them has an associated hard X-ray (2--10~keV) point source.
We suggest that the two symmetric radio bars can be explained by two 
precessing jets ejected from an AGN.
Beyond the central regions, we find two extended X-ray sources to the
southwest and northeast of the cluster center which are apparently
associated with merging subclusters.
The main Abell~2626 cluster and these two subclusters are extended along
the direction of the Perseus-Pegasus supercluster, and we suggest that 
Abell~2626 is preferentially accreting subclusters and groups from this 
large-scale structure filament.
We also find an extended X-ray source associated with
the cluster S0 galaxy IC~5337; the morphology of this source suggests that
it is infalling from the west, and is not associated with the southwest
subcluster, as had been previously suggested.
\end{abstract} 


\section{Introduction}
\label{sec:intro}

The central regions of clusters of galaxies are among the most interesting 
and physically active areas in the Universe.
In many clusters of galaxies, there is a central 
peak in X-ray surface brightness, with a central cooling time which is less 
than the age of the cluster \citep[e.g.,][]{CRB+07}.
In the classical cooling flow picture, the 
rapid cooling of gas in this region leads to the loss of central 
pressure \citep{Fab94}.
If there is no additional heating, the cooled gas would flow into the center
in a cooling flow.
The gas should continue to cool to very low temperatures.
However, recent high resolution X-ray spectra from {\it 
XMM-Newton} and {\it Chandra} show that there is not enough cooler X-ray gas 
to be consistent with a classical cooling flow \citep{Pet+01,PKP+03, 
PF06}. 

The lack of sufficient cooler materials probably means that
there is an additional heating source within the central cooling core 
region. 
Very often, the clusters host central dominant galaxies
which are also strong radio sources. 
In some of the clusters, there is strong evidence that the central radio 
sources are interacting with the surrounding X-ray emitting plasma. 
The radio sources blow ``bubbles'' in the X-ray emitting gas, displacing 
the gas and compressing it into shells around the radio lobes. At the same 
time, the radio sources are confined by the X-ray emitting gas 
\citep{Fab+00, BSM+01}.
In some clusters, 
``ghost bubbles'' at larger radii are seen which are weak in radio 
emission except at low 
frequencies \citep{Mcn+01, CSB+05}.
They are believed to be  previous eruptions of the radio sources, which are
episodic.
In some cases, buoyantly rising bubbles may entrain cooler X-ray 
gas from the centers of the cooling cores
\citep{CBK+01, FSK+02, FSR+04}.
Some radio sources previously 
classified as cluster merger radio relics may actually be displaced radio 
bubbles from the central radio sources
\citep{FSK+02, FSR+04}.
In some recent studies, there is even evidence of radio jets restarted at 
a different position angle from a previous outburst, which may be 
due to
a precessing supermassive blackhole \citep{GFS06}.
With high resolution 3D simulations, \citet{VR06} have shown that multiple
bursts from the AGN along the same direction fail to produce a long-term
balance of heating and cooling.
The reason is that a low-density channel is evacuated by a previous 
jet, while the subsequent jets can pass through the channel freely 
without heating the cooling core region. 
It has been suggested that jet precession is a possible solution to this 
problem.
Alternatively, turbulence, rotation, or bulk motions in the 
intracluster medium (ICM) might 
also move or disrupt the channel made by the radio source, and keep the 
energy from just going down the same channel \citep[e.g.,][]{HBY+06}.

Abell~2626 \citep[$z=0.0573$,][]{ACO89} hosts a moderate cooling flow 
of
$\sim 50 \, M_{\odot}$ yr$^{-1}$ based on {\it Einstein} data
\citep{WJF97}.
A double nuclei cD 
galaxy is sitting at the center of the cluster.
From high resolution VLA images, \citet{GBF+04} recognized that there is  
an unresolved radio core coincident with the central cD galaxy.  There is 
a 
small jet-like feature extended towards the southwest direction from 
the unresolved radio core.  In addition, 
there are two unusual amorphous, symmetric radio ``bars'' running parallel
at opposite sides of the central cD galaxy.  These compact features are 
distinct from and embedded in diffuse extended radio emission 
(a mini-halo).
\citet{GBF+04} believed that the radio bars may 
represent an earlier evolutionary stage of jets injected by the central 
source.   
Thus, Abell~2626 is a good candidate for studying the interaction 
between the X-ray emitting intracluster medium and the radio emitting 
plasma \citep{RLB+00, GBF+04, MOE04}.  
Previous {\it ROSAT} X-ray and VLA radio observations suggested that there is
an X-ray excess which is spatially correlated with the radio source, but 
failed to find strong X-ray deficits (holes) in Abell~2626 
\citep{RLB+00}.  \citet{RLB+00} carried out numerical simulations, and 
showed that even a weak jet (internal Mach number = 5) 
should produce a significant hole along the line-of-sight to the radio jet.  
They suggested that the lack of X-ray holes may be due to mixing of
thermal and radio plasma in the region, although their 
numerical simulations did not distinguish between these two components.  
They also ran simulations in which the jet turned off, which resembled
the observations of weak X-ray holes.
Another possibility they noted is the imprecise model they used 
to subtract out from the symmetric X-ray emission to obtain a residual image.  
They searched for X-ray and radio correlations from the residual image
but failed to find strong X-ray deficits (holes) displaced by the radio 
plasma.  With the excellent spatial resolution provided by {\it 
Chandra} and deeper observations provided by both {\it Chandra} and {\it 
XMM-Newton}, we are able to study the complicated central region of 
Abell~2626 in more detail.

While many radio sources are associated with AGNs in cluster galaxies,
there exist other types of extended radio emission associated with the
diffuse intracluster medium without clear galaxy hosts.
These extended radio sources can be divided into cluster-wide halos, 
relics, and mini-halos \citep{FG96, FG07}.
Abell~2626 contains a candidate mini-halo \citep{GBF+04}.
Cluster-wide halos and 
relics are large in scale ($\sim 1$ Mpc), with halos located 
in the centers of clusters, while the relics are generally located at the 
periphery.
Very often, these two types of large scale 
diffuse radio emission are found in clusters without cooling cores.
This indicates a possible origin in cluster mergers, which are believed
to disrupt cooling cores and to re-accelerate radio-emitting particles 
\citep{Bru04, BGB07}.
However, 
some cooling core clusters with dominant radio galaxies host 
mini-halos at the cluster center with extended radio emission up to 
0.5~Mpc.
The morphology of the mini-halos is different from that of the
radio bubbles at the centers of many cool core clusters, which
show a strong X-ray/radio interaction in which the radio 
lobes displace X-ray gas.
This hints at a different transport mechanism (not jets)
to explain the extended property of the mini-halos.  
The typical time scales for radiative losses by relativistic electrons 
($\sim 10^7-10^8$ yr) 
in cooling cores with strong magnetic fields
are much shorter than the diffusion times ($\sim 10^9$ yr) for the electrons
to be transported from the center of the cooling core out to a fraction of 
a Mpc.  Hence, the diffuse radio emission from mini-halos suggests that 
an {\it in-situ} re-acceleration mechanism is needed for the electrons.
The origin of the mini-halos has remained unclear, 
but it has 
been suggested that the interaction of the central radio galaxy and the 
intracluster medium (re-)accelerates particles to 
relativistic energies.
Recently, \citet{GBF+04} modeled
Abell~2626 with an electron re-acceleration model driven by MHD turbulence 
amplified by the compression of the magnetic field in the cooling core 
region.  
They have shown that the model agrees with 
various observational constraints such as the observed radio brightness 
profile, the integrated synchrotron spectrum, and the radial steepening of
the radio spectrum. 
They conclude that only a tiny 
fraction ($\sim 0.7\%$) of the maximum power that could be extracted from 
the cooling core (based on the standard cooling flow model) is needed for 
re-acceleration.
Since the re-acceleration model for the origin of the mini-halo assumes a 
physical connection between thermal and relativistic plasma, it is 
essential to perform new X-ray observation to derive accurately the 
properties of the ICM (e.g., cooling radius, mass accretion rate, as well 
as morphology of the X-ray emission).   On another hand, \citet{PE04} 
suggested that the re-acceleration is of hadronic origin which should be 
best tested by $\gamma$-ray observations.

In this paper, we present a detailed study of Abell~2626 from {\it 
XMM-Newton} and {\it Chandra} observations focused on the X-ray and radio 
interaction.
The paper is organized as follows.
In \S~\ref{sec:data}, we describe observations and data reduction.
The X-ray images are analyzed in \S~\ref{sec:image}, with an emphasis on 
the X-ray morphology of the overall region, the central region, and the S0 
galaxy IC~5337 in \S~\ref{sec:image_morph}.  In particular, two extended 
X-ray emission regions possibly associated with subclusters are discussed in 
\S~\ref{sec:image_morph_outer}.
The point sources detected in the X-ray data are discussed in 
\S~\ref{sec:sources_detection}.
The X-ray surface brightness profile is studied in \S~\ref{sec:SB}.  
Various substructures identified on residual 
maps are presented in \S~\ref{sec:residual}.
Spectral analysis is presented in \S~\ref{sec:spectrum}.  
In particular, cluster profiles, mass deposition rate, cooling and thermal 
conduction time scales, hardness ratio maps, 
possible AGNs associated with the cD galaxy IC~5338 and the S0 galaxy 
IC~5337 are discussed  in this section.
Throughout the data analysis sections (\S\S~\ref{sec:image} \& 
\ref{sec:spectrum}), some discussions related to the interpretation of the 
results are presented to help readers understand the data.  We then 
discuss the kinematics of the S0 galaxy and the cooling flow interaction 
with the central radio source in \S\S~\ref{sec:discuss_S0} \& 
\ref{sec:discuss_radio} in detail, respectively.
We summarize our work in 
\S~\ref{sec:conclusion}.
We assume $H_0 = 71~{\rm km~s}^{-1}{\rm ~Mpc}^{-1}$, $\Omega_M = 0.27$ and
$\Omega_{\Lambda} = 0.73$ throughout the paper. At a redshift $z =
0.0573$, the luminosity distance of Abell 2626 is 252.8~Mpc,
and $1\arcsec = 1.10$~kpc.
The virial radius is approximately 1.53 Mpc \citep{MFS+98} for the average
cluster temperature of $T_X = 3.0$ keV \citep{WJF97}.

\section{Observations and Data Reduction}
\label{sec:data}

\subsection{XMM-Newton Data}
\label{sec:data_xmm}
 
Abell~2626 was observed (Obs.\ ID 0148310101) on 2002 December 28
by {\it XMM-Newton} for $\sim 38$ and $\sim 42$ ks with the EPIC PN and 
MOS cameras, respectively.
Extended Full Frame mode (PN) and Full Frame mode (MOS) were used with 
thin optical filters. 
{\it XMM-Newton} data reduction was done with SAS version 5.4.1. All 
the data were reprocessed using the tasks {\it epchain} and {\it emchain}. 
We included only events with FLAG=0 in our analysis.
We chose only events with PATTERN = 0-4 (0-12) for PN (MOS) data.

Background flares were rejected by performing a $2 \sigma$ clipping of the 
100~s binned light curves 
in the 12-14 keV and 10-12 keV band 
for the PN and MOS cameras, respectively, 
which yielded the out-of-time corrected exposure times of 35,563~s, 40,300~s
and 40,600~s of cleaned data for the PN, MOS1 and MOS2 cameras,
respectively.
Blank-sky background files collected by 
D.~Lumb\footnote{ftp://xmm.vilspa.esa.es/pub/ccf/constituents/extras/background/} 
were used during the analysis.
The observed background levels, estimated from the total count rate in the 
whole field of view in the hard band (10 -- 12~keV and 12 -- 14~keV for 
MOS and PN, respectively), are 1.07, 0.99 and 1.26 times those in the 
blank-sky backgrounds for MOS1, MOS2 and PN, respectively.  We have 
renormalized the blank-sky backgrounds accordingly.  
The uncertainty of the PN background is larger than that for the MOS
detectors.  The effect of the larger PN background uncertainty and some 
other background uncertainties on the 
spectral analysis are addressed in \S~\ref{sec:spectrum:background}.
The image and hardness ratio map analyses were done using the MOS data 
only (\S\S~\ref{sec:image} \& \ref{sec:spectrum:substructures}), and we 
checked that the results of the analysis would not be 
affected by a change in the background normalization of $\pm$10\%.
The vignetting effect for the image and hardness ratio 
map analysis has been corrected using the exposure maps.

\subsection{Chandra Data} 
\label{sec:data_chandra}

A {\it Chandra} observation (Obs.\ ID 3192) of $\sim 25$ ks  was made
on 2003 January 22 in Very Faint mode.
The center of the cluster was positioned 3\arcmin\ from the edge
of the S3 chip (midway between the node boundaries).
The focal plane temperature was $-120$ C.

The {\it Chandra} data reduction was done with CIAO version 3.1, using
calibration products from CALDB~2.28. 
The latest gain file (CALDB GAIN 2.28) and geometry file (CALDB~2.9) at 
the time of reduction were used.
The reduction was done in Very Faint mode to remove more particle background.
Background flares were rejected by performing the standard clipping using 
the {\it lc\_clean} routine written by 
M.~Markevitch\footnote{http://cxc.harvard.edu/contrib/maxim/acisbg/}.
This yielded 22,959~s of cleaned data.
Since the whole field-of-view is dominated by the cluster X-ray emission, 
internal background cannot be used.
The blank-sky background files collected by M.~Markevitch were used 
during the analysis.
The difference between the observed and the blank-sky high energy 
(9--12 keV) background count rates
was less than 1\%, and this small difference was removed by renormalizing
the background.
The influence of the background on the spectral fitted 
measurements (temperature profile) has been checked by 
varying the background normalizations by $\pm 10\%$, and the results are 
not significantly changed.
To check if there might be a significant difference 
between the Galactic background in the source and blank-sky observations, 
another thermal component with a fixed low temperature of 0.2~keV has 
been added \citep{Mar+03, LWP+02}.  Normalization of the additional 
component has been allowed to 
be negative for the case of more Galactic emission in the blank-sky 
observations compared to the cluster observation.  All the results are 
basically unchanged.  Fixing the temperature of the additional component 
to be the outermost temperature that can be determined from the {\it 
Chandra} field of view also gives essentially the 
same 
results.  We conclude that the {\it Chandra} results are insensitive to 
uncertainties in the background.

\section{X-Ray Image Analysis}
\label{sec:image}

\subsection{X-ray Morphology}
\label{sec:image_morph}

\subsubsection{Global morphology}
\label{sec:image_morph_outer}

\begin{figure}
\includegraphics[angle=0,width=8.2cm]{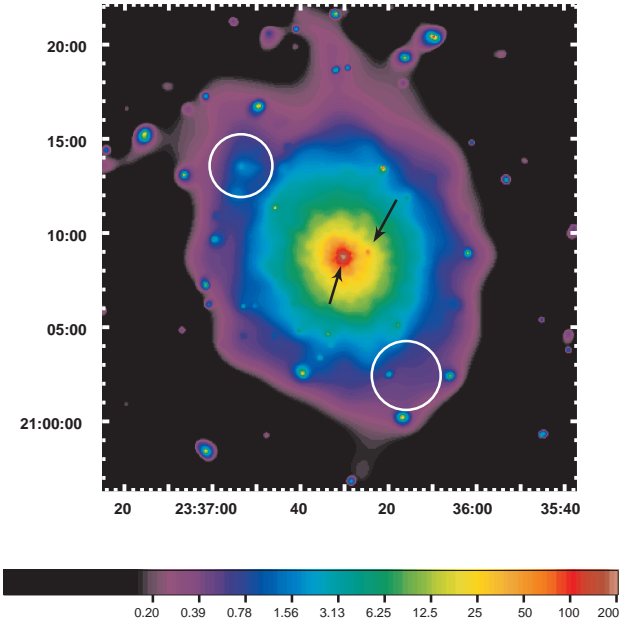}
\caption{
Background-subtracted, exposure-corrected, adaptively smoothed
mosaic of the {\it XMM-Newton} EPIC MOS1 and MOS2 images
in the 0.3--10 keV band.
The image was smoothed to a signal-to-noise ratio of 3.
The units for the color scale are counts per binned pixel, where each 
binned pixel has a size of $4\farcs1\times4\farcs1$.
The arrows on the left and right indicate the cD galaxy IC~5338 and the S0 
galaxy IC~5337, respectively.
Two outer extended emission regions are indicated with white circles.
It should be noted that there is a chip gap at about $5\farcm5$ from the 
center which makes the northeast extended X-ray source appear to be more 
distinct than it actually is. 
}
\label{fig:smoothed_imageXMM}
\end{figure}

The background subtracted, exposure-corrected, adaptively smoothed 
mosaic 
of the {\it XMM-Newton} EPIC MOS1 and MOS2 images is shown in 
Figure~\ref{fig:smoothed_imageXMM}.
The PN image was not included because of its different spectral response
and because the chip gaps and bad columns produced cosmetic artifacts
in the smoothed image.
The image was smoothed to a signal-to-noise ratio of 3 per smoothing beam.

The global X-ray image of Abell~2626 appears to be roughly azimuthally 
symmetric.
The brightest region at the center corresponds to the position of the cD
galaxy IC~5338. 
There is a source associated with the cluster S0 galaxy IC~5337 which is
1\farcm3 west of the center of the cluster.
Excess X-ray emission is seen to the southwest of the cluster, extending
from $\sim$1\arcmin\ out to $\sim$7\arcmin\ (see \S~\ref{sec:residual}).
\citet{MGW96} identified a subcluster to the southwest of this region 
based on optical 
observations of the 
galaxy population.  The subcluster identified is centered roughly at the 
southwest edge of the {\it XMM-Newton} field of view, but the distribution 
of the galaxies covers a region where the excess X-ray emission is seen.
\citet{MGW96} show that the subcluster is bound to and falling into
the main cluster.
Our {\it XMM-Newton} image shows, for the first time, extended X-ray 
emission possibly associated with the subcluster in the southwest 
direction.
The  extended X-ray emission can be viewed on a background subtracted, 
exposure corrected, adaptively smoothed image.
The location of the extended emission possibly associated with the  
subcluster is circled at the southwest corner in 
Figure~\ref{fig:smoothed_imageXMM}.
The extended emission at the southwest corner is not obvious,
but the slight deviation from spherical symmetry in that region can be noted.
The shape of the extended emission region is difficult to characterize
due to the low signal-to-noise ratio near the edge of the cluster.
By inspecting various images, we chose a circular region centered on the
highest excess X-ray emission, with a radius chosen to be small enough so
that the region did not overlap with the nearly circular cluster emission
of Abell~2626.  Our choice for this region is justified by the hardness
ratio map below (Figure~\ref{fig:HRmap_xmm_large}) which shows that the
extended region chosen appears to be harder than its surrounding.
Compared to the X-ray emission at the same 
radius from the main cluster center (the cD galaxy IC~5338) with all 
point 
sources and other extended emission excluded, 
the excess emission is significant at the $4.6\, \sigma$ level.
Our residual images (\S~\ref{sec:residual} below)
also indicate the possible signatures of a merger.

Another extended X-ray source can be seen about 7\arcmin\  northeast
of the center of Abell~2626 (Figure~\ref{fig:smoothed_imageXMM}).
The criteria used to define the northeast region are the same as the 
southwest one.
This feature is significant at the $11.6\, \sigma$ level.
It should be noted that there is a chip gap at about 5\farcm5 from
the center which makes the extended X-ray source appear to be
more distinct than it actually is.
No cluster or group structure has been identified previously in this region.
However, Abell~2626 is known to be associated with the
Perseus-Pegasus supercluster, a filament of clusters of galaxies extending
for as much as $\sim$300 Mpc
\citep{BB85a,BB85b,ZZS+93,EET+01}.
The southwest to northeast extension to the structure in our X-ray image
around Abell~2626 is elongated along the Perseus-Pegasus filament.
This may indicate that the cluster is preferentially accreting subclusters
and groups from this supercluster.

\subsubsection{Central Region}
\label{sec:image_morph_cD}

The background subtracted, exposure corrected, adaptively smoothed
{\it Chandra} image of the central 3\farcm5$\times$2\farcm4\
is shown in Figure~\ref{fig:smoothed_imageChandra} (left panel). 
The image was smoothed to a signal-to-noise ratio of 3.
The brightest region of the X-ray emission in the cluster is centered on
the southwest nucleus of the central cD galaxy.
This source is possibly extended in both soft (0.3--2 keV) and full 
band (0.3--10 keV) images when detected by {\it wavedetect} in CIAO, but 
nothing was found by {\it wavedetect} in the hard band (2--10 keV).
However, examination of the raw 
soft band (0.3--2 keV) image shows two peaks superposed on the two optical 
nuclei of the central cD galaxy (lower left panel of 
Figure~\ref{fig:2nuclei}).
There is also a hard point source seen in the raw 
hard band (2--10 keV) image (lower right panel of 
Figure~\ref{fig:2nuclei}).
The position of this hard point source agrees with the southwest cD 
nucleus.
The hard point source is separated from the peak of the  
radio core by less than 1\arcsec.
This source can also be seen 
in the hardness ratio map (Figure~\ref{fig:HRmap_chandra_halo} below).
The hard point source and its likely association with the AGN at the
center of the southwest cD nucleus and radio core are discussed further
in \S~\ref{sec:AGN}.
This source is located along an arc of X-ray emission 
extending from the northeast to the west to the south of the cluster 
center.

Previous {\it ROSAT} X-ray and VLA radio observations suggest that there 
is enhanced X-ray emission spatially coincident with the radio source 
\citep{RLB+00}. 
In Figure~\ref{fig:smoothed_imageChandra} (left panel), the contours 
from the 
VLA 1.5 GHz B-array radio image are shown in green
\citep{GBF+04}.
The position of the radio core agrees with the brightest region of
X-ray emission.
In particular, the radio core is centered at the southwest cD 
nucleus rather than the northeast one.
There is an arc of X-ray emission which may correspond to the curved radio
jet feature to the south of the core.

However, the most unusual features of the radio image are the two elongated
radio bars to the north and south of the radio core.
We find there is no obvious correlation between the two radio bars
and the {\it Chandra} X-ray image in
Figure~\ref{fig:smoothed_imageChandra} (left panel).
No obvious X-ray deficit (holes) is found on the {\it Chandra} X-ray 
image, which is consistent with the finding of \citet{RLB+00}.
We have generated residual maps to further investigate our finding, and 
the results will be presented in \S~\ref{sec:residual}.
Our interpretation is that the radio bars are thin tubes
(see \S~\ref{sec:discuss_radio} below).

One interesting feature is that there is an X-ray excess ``tongue'' which 
extends from the 
central cD galaxy to the southern radio bar, and there is a
similar but weaker tongue in the northern direction as well
(upper right panel of Figure~\ref{fig:2nuclei}).
The tongues are best seen in the residual map in
Figure~\ref{fig:residual_chandra_halo} of \S~\ref{sec:residual}.

Figure~\ref{fig:smoothed_imageChandra} (left panel)
also shows radio contours (white) from the VLA 1.5 GHz 
C-array \citep{GBF+04} overlaid on the {\it Chandra} image showing the 
radio mini-halo in Abell~2626.
The mini-halo is roughly confined within 60\arcsec\ with a diamond shape.  
The distance from the center to the NE or SW corners of the diamond is 
about 50\arcsec, while to the NW or SE corners it is about 70\arcsec.
A discussion on the mini-halo will be presented in 
\S~\ref{sec:discuss_radio}.

\subsubsection{The S0 Radio Galaxy IC 5337}
\label{sec:image_morph_S0}

\begin{figure}
\includegraphics[angle=0,width=5in]{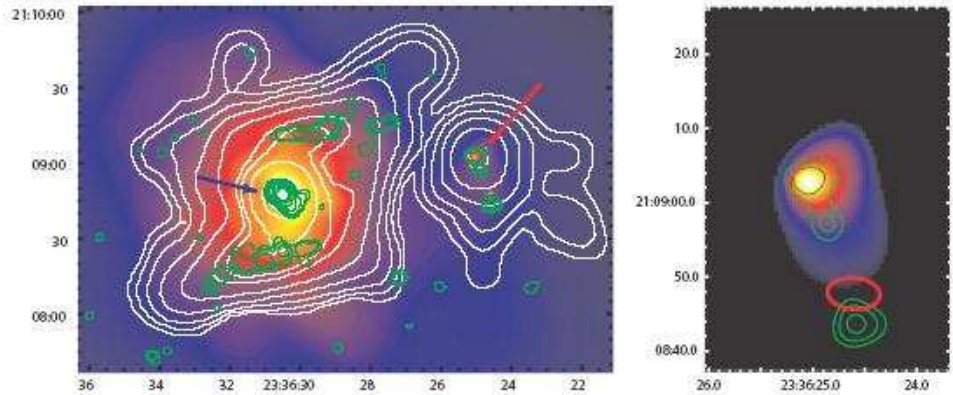}
\caption{
{\it Left panel:}
Background subtracted, exposure-corrected, adaptively smoothed
{\it Chandra} image of the central 3\farcm5$\times$2\farcm4 
region of the cluster
in the 0.3--10 keV band.
The image was smoothed to a signal-to-noise ratio of 3.
The {\it Chandra} image shows the details of the inner region of the cooling 
core.
The white solid lines are 1.5 GHz C-array radio contours showing the 
mini-halo.
The green solid lines are 1.5 GHz B-array radio contours showing the
radio bar structures.
Both radio contours were taken from \citet{GBF+04}.
The arrows on the left and right indicate the cD galaxy IC~5338 and the S0 
galaxy IC~5337, respectively.
{\it Right panel:}
Background subtracted, exposure-corrected, adaptively
smoothed {\it Chandra} image centered on the S0 galaxy IC~5337.
The image was smoothed to a signal-to-noise ratio of 4,   
with an intensity scale chosen to show the bow-shock-like shape.
The southern 
source (red ellipse) was removed and replaced by the average surrounding 
X-ray intensity before adaptively smoothing.
The green contours are from the VLA 1.5 GHz B-array radio image taken from  
\citet{GBF+04}, with the lowest contour level being a signal-to-noise 
ratio of 3.
}
\label{fig:smoothed_imageChandra}
\label{fig:S0}
\end{figure}

\begin{figure}
\includegraphics[angle=0,width=5in]{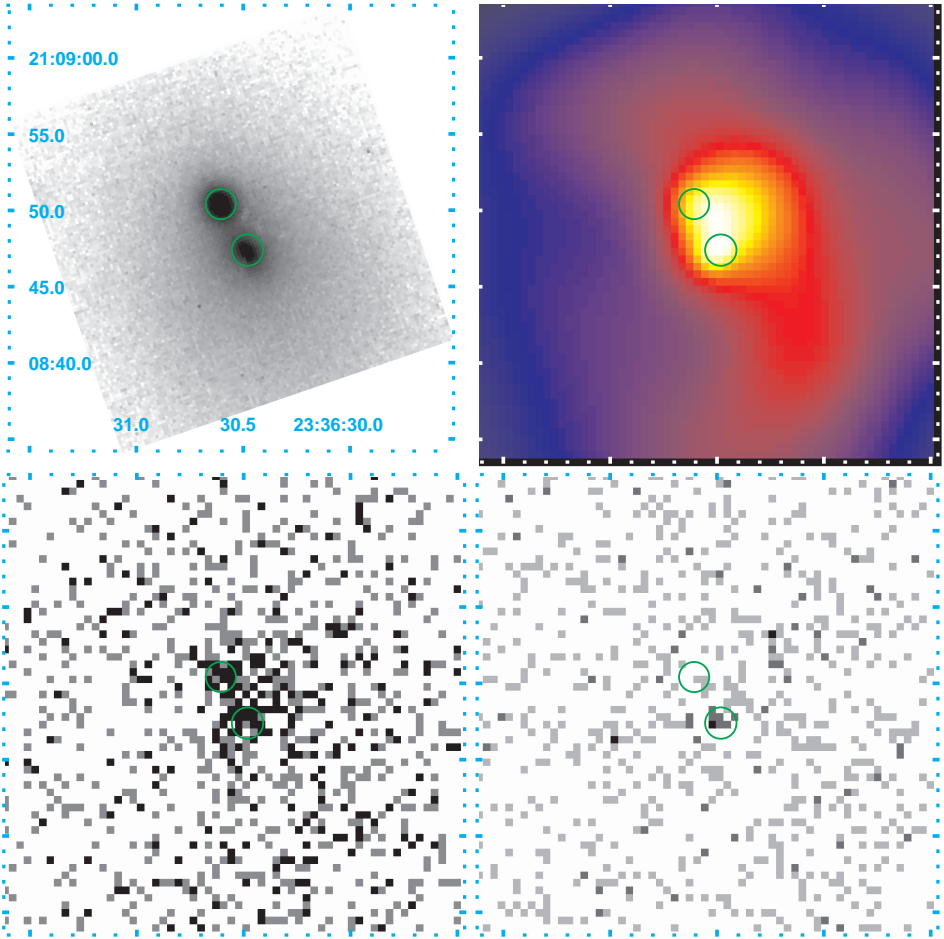}
\caption{
Images of the two nuclei of the cD galaxy IC~5338 in Abell~2626.
All images show the same field of view of $30\arcsec\times30\arcsec$.
{\it Upper left panel:} Optical image from {\it HST} archive.
The image was taken with WFPC2 using the F555W filter.
The two green circles, included in the other panels, are centered at the two 
optical nuclei observed by {\it HST}.
{\it Upper right panel:} Background-subtracted, exposure-corrected, 
adaptively smoothed {\it Chandra} image in 0.3--10~keV band.  The image 
was smoothed to a signal-to-noise ratio of 3.  The color represents the 
X-ray intensity from high (white yellow) to low (dark blue).
{\it Lower left panel:} Raw {\it Chandra} image in the soft (0.3--2~keV) band.
Two intensity peaks can be identified.
{\it Lower right panel:} Raw {\it Chandra} image in the hard 
(2--10~keV) band.
Only the southwest nucleus corresponds to the peak in the hard band.
}
\label{fig:2nuclei}
\end{figure}

About 1\farcm3 west of the central cD galaxy, there is an X-ray source 
associated
with the S0 galaxy IC~5337
(left panel of Figure~\ref{fig:smoothed_imageChandra}).
This source is clearly extended in the soft band image, but a hard band
image shows that there is a point source coincident with the nucleus of
IC~5337.  The possibility that the S0 nucleus is an AGN will be 
discussed below (\S~\ref{sec:AGN}).
To the south of IC~5337, there is another X-ray point source which is
not associated with any known galaxy.
As indicated in Figure~\ref{fig:smoothed_imageChandra}, 
there is a
nearby radio source which may be associated with this X-ray source.

The right panel of Figure~\ref{fig:S0} shows the bow-shock-like shape of 
the extended
X-ray emission associated with the S0 galaxy IC~5337.
The image is created by removing and replacing the southern point source 
by the average surrounding X-ray intensity, and then adaptively smoothing 
to a signal-to-noise ratio of 4.

The S0 galaxy IC~5337 also is a radio source
(Figure~\ref{fig:smoothed_imageChandra}).
Radio contours \citep{GBF+04} are overlaid in Figure~\ref{fig:S0} (right 
panel).
There is a component centered on the S0 galaxy, and a 
component located at a position of about 190\arcdeg\ (measured from 
north to east) of the S0 galaxy on the 1.5 GHz B-array map.
The 1.5 GHz C-array map shows three radio tails in the south, southwest 
and west directions.
The kinematics of the S0 galaxy will be discussed in 
\S~\ref{sec:discuss_S0}.

\subsection{Source Detection}
\label{sec:sources_detection}

\begin{figure}
\includegraphics[angle=0,width=8.2cm]{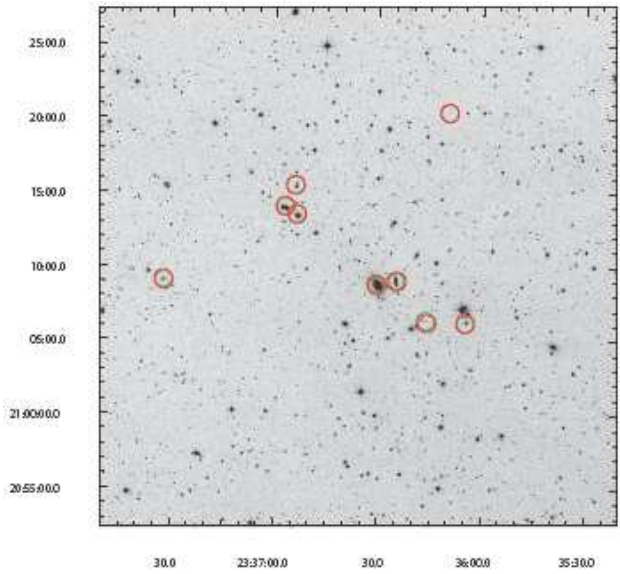}
\caption{
Optical image from the Digital Sky Survey showing
the 9 {\it XMM-Newton} X-ray sources with identifications in NED
(Table~\ref{table:source}).
The red circles are centered on the X-ray positions determined with
{\it XMM-Newton}.
There are green circles centered on the optical positions from 
NED, but they overlap so completely with the red circles that they are
barely visible.
NED objects are considered possible identifications if they are
within
20\arcsec\ of an X-ray source.
In fact all optical counterparts identified are within 11\arcsec.
}
\label{fig:optical}
\end{figure}

We used the {\it XMM} SAS task {\it eboxdetect} to detect a total
of 90 distinct point sources on the {\it XMM} images;
the detections were done separately on each of the three detectors,
and then the source lists were combined.
The sources were confirmed by inspection, 
and some low-level detections at the edge of the FOV or bad pixel gaps 
were removed.
Here we discuss only the sources with possible optical counterparts
within 20\arcsec.
In fact all optical counterparts identified are within 11\arcsec.
We adopt the NED source with an accurate position which is closest to
the X-ray position,  
The possible associations of X-ray point sources with AGNs in the central cD
galaxy IC~5338 and the S0 galaxy IC~5337 were noted above
(\S\S~\ref{sec:image_morph_cD} \& \ref{sec:image_morph_S0}), and will
be discussed in more detail below (\S~\ref{sec:AGN}).
Nine X-ray sources are included in the list (Table~\ref{table:source}).
Figure~\ref{fig:optical} is an optical image from the 
Digital Sky Survey with the 9 X-ray sources marked.
Two of these are the central cD galaxy  IC~5338 and the S0 galaxy IC~5337
discussed above. 
Interestingly, three of the X-ray sources correspond to galaxies which are 
located in the region of extended X-ray emission about 7\arcmin\ northeast
of the center of Abell~2626 (Figure~\ref{fig:smoothed_imageXMM}).
However, one of these possible optical IDs (2MASX J23365722+2114032) has a 
redshift of 0.038, and thus is not directly associated with 
Abell 2626.

\begin{figure}
\includegraphics[angle=0,width=8.2cm]{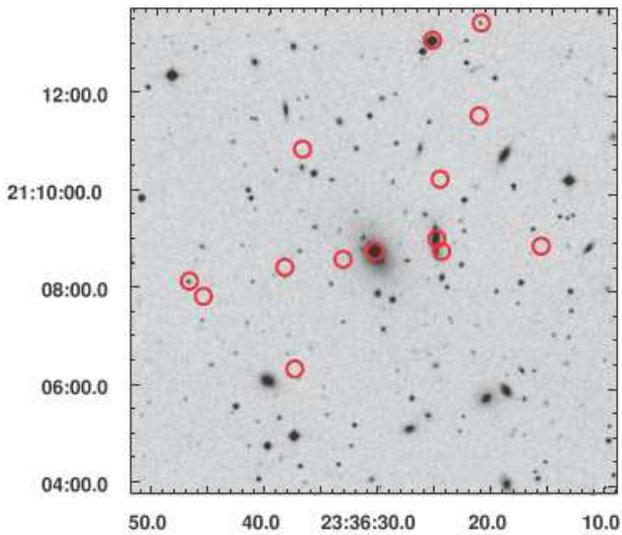}
\caption{
Optical image from the Digital Sky Survey showing the positions
(red circles) of the 14 X-ray point sources detected with {\it Chandra}
(Table~\ref{table:source_chandra}).
}
\label{fig:optical_chandra}
\end{figure}

On the {\it Chandra} image, 
14 point sources were detected by
{\it wavedetect} in CIAO using the default settings.
Of these, only the central cD IC~5338 galaxy and the S0 galaxy IC~5337
have possible identifications in NED with positions that agree
with the X-ray to within 10\arcsec.
Table~\ref{table:source_chandra} lists the positions of the
14 {\it Chandra} X-ray point sources.
Figure~\ref{fig:optical_chandra} is an optical
image from the Digital Sky Survey with the
locations of the
14 X-ray sources marked.

Unless specified, all the sources detected are excluded in the analysis of 
extended emissions in the subsequent sections.

\subsection{Surface Brightness Profile}
\label{sec:SB}

The azimuthally-averaged X-ray surface brightness profile was extracted
separately for the {\it Chandra} and {\it XMM-Newton} data.
Except for the two extended X-ray clumps to the NE and SW and the structure
in the very center, Abell~2626 appears to be a nearly relaxed cluster.
The profiles were fitted using a double beta model profile of the form
\begin{equation} \label{eq:twobeta}
S(r)
=
\sum_{i=1,2}
S_i(0)
\left[ 1 + \left( \frac{r}{r_{ci}} \right)^2 \right]^{-3 \beta_i+ 1/2}
\, ,
\end{equation}
which has been shown to provide a good fit to relaxed clusters with cooling
cores \citep{XW00}.

\begin{figure}
\includegraphics[angle=0,width=8.2cm]{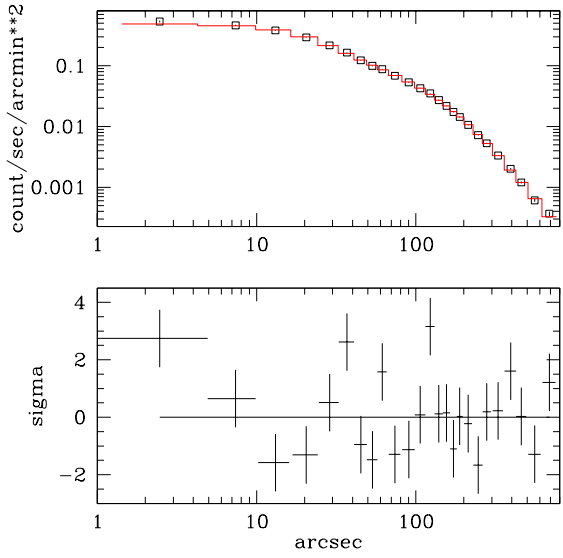}
\caption{
{\it Upper panel}: Azimuthally averaged X-ray surface brightness
profile from the {\it XMM-Newton} MOS1 and MOS2 observations of Abell~2626.
The squares are the X-ray data, while the
solid line is the fit to a double $\beta$ model.
The tiny error bars within the squares on the surface brightness are too 
small to be easily seen.
{\it Lower panel}:
The residuals to the best fit double beta model, 
in unit of $\sigma$.
The horizontal error bars show the widths of the bins used to accumulate
the counts.
The vertical error bars are at the 1 $\sigma$ level.
}
\label{fig:surface_brightness_xmm}
\end{figure}

The background subtracted and exposure corrected surface brightness 
profile from the {\it XMM-Newton} MOS1 and MOS2 data is shown in
Figure~\ref{fig:surface_brightness_xmm}. 
The PN data were not included since the spectral response is different 
which complicates the deprojection of the electron density profile 
below (\S~\ref{sec:density}).  Also, the fitted surface brightness model
was used to 
create residual images (\S~\ref{sec:residual} below), and including the 
PN image would produce cosmetic artifacts due to the larger chip gaps and 
greater number of bad columns in the PN camera.
All of the X-ray point sources except the central cD galaxy were excluded 
from 
the data.
The extended X-ray emission around the S0 galaxy was also excluded, as
it produced a feature in the cluster surface brightness profile.
We also excluded the extended X-ray emission regions to the NE and SW
of the cluster which were described in and the criteria of defining the 
regions were explained in  
\S~\ref{sec:image_morph_outer}.
The annuli were chosen such that all the widths are larger than the 
FWHM of the PSF at that radius.
With this choice, all the bins contain at 
least 1000 counts after background subtraction for the {\it XMM-Newton} 
MOS1 and MOS2 data.
With these very good statistics, the error bars on the surface brightness
are too small to be easily seen in the top panel of
Figure~\ref{fig:surface_brightness_xmm};
instead, we present the residuals in the lower panel.
The parameters of the best-fit double beta model are
$S_1 (0) = 0.40^{+0.02}_{-0.02}$ and
$S_2(0) = 0.10^{+0.01}_{-0.01}$ counts~s$^{-1}$~arcmin$^{-2}$,
$r_1 = 34.8^{+7.1}_{-5.6}$ and
$r_2 = 128.5^{+8.1}_{-7.2}$ arcsec,
and
$\beta_1 = 0.91^{+0.26}_{-0.18}$ and
$\beta_2 = 0.73^{+0.02}_{-0.01}$.
Visually, the double beta model fit seems adequate.
However, with the very small errors on the measured surface brightness,
the fit is actually not very good, with a reduced $\chi^2$ of 2.3.
In addition to the excess at the center, associated with the central
cD galaxy and radio source, within 200\arcsec\ there are several other
regions of excess or deficient emission compared to the model.
These features are reproduced (albeit with poorer statistics) in
the {\it Chandra} surface brightness data, and in the data from
the {\it XMM-Newton} PN detector.
These residuals may indicate that the central region is disturbed
by effects of the central radio source beyond the current extent of
the source,
or that the cluster is not completely relaxed due to a merger.

\subsection{Residual Images}
\label{sec:residual}

\begin{figure}
\includegraphics[angle=0,width=5in]{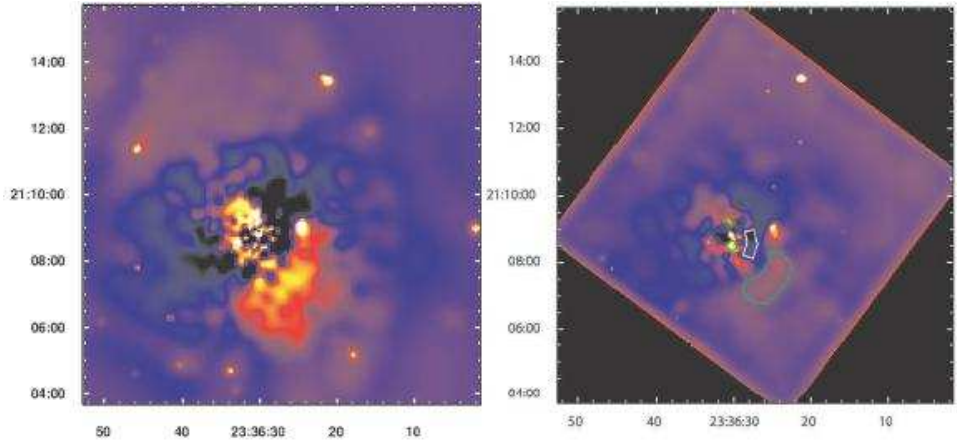}
\caption{
{\it Left panel:}
Residual map from the {\it XMM-Newton} MOS1 and MOS2 observations of
Abell~2626,
generated from a background-subtracted image.
Yellow represents the largest excess, while black is the
greatest deficit.
{\it Right panel:}
Residual map from the {\it Chandra} observation of Abell~2626,
generated from a background-subtracted image.
Yellow represents the largest excess, while black is the
greatest deficit.
There is an excess coincident with the center of the central cD galaxy,
which is partly surrounded by a trough of reduced X-ray surface
brightness at a radius of about 15\arcsec.
There are two localized regions of excess south and north of the
central peak, which are indicated with green circles.
There is an extended deficit about 50\arcsec\ west of the center, which
is indicated with a white polygon.
An extended region of excess emission is seen about 110\arcsec\ SW of the
center, which is indicated with a green polygon.
\vskip 15mm
}
\label{fig:residual_xmm}
\label{fig:residual_chandra}
\end{figure}

To better understand the origin of the residuals in the surface brightness 
fit
and the dynamics of the intracluster gas, we created residual
maps for both {\it XMM-Newton} and {\it Chandra} images
(Figures~\ref{fig:residual_xmm} and \ref{fig:residual_chandra_halo}).
Simulated azimuthally-symmetric X-ray images were produced which exactly
followed the double beta
model fits to the surface brightness.
These simulated images were subtracted from the adaptively smoothed images.
Figure~\ref{fig:residual_xmm} shows the
{\it XMM-Newton} and {\it Chandra} residual images with the same field of 
view.
Both residual
maps show similar structures;
the {\it XMM-Newton} image has great contrast due to the better statistics,
while the {\it Chandra} image shows more detailed structures.
Substructures can be identified with many of the residuals seen in the
maps.
A large region of excess emission can be seen about
110\arcsec\ southwest of the cD galaxy, which corresponds to the 
peak of the surface brightness residuals around that radius. Other regions
of excess and deficit within 100\arcsec\ can also be identified from the  
residual map of the {\it Chandra} image 
(right panel of Figure~\ref{fig:residual_chandra}).
In the surface brightness residuals,
the central excess corresponds to the central
bright point in Figure~\ref{fig:residual_chandra} (right panel).
The trough around 15\arcsec\ corresponds to the dark ring around the
central excess in the residual map which can be seen more clearly on 
Figure~\ref{fig:residual_chandra_halo}.
The peak at around 30\arcsec\  is probably due to the two bright regions
due north and south in the residual map
(labeled as two green circles).
The trough around 50\arcsec\ corresponds to the dark region indicated by
a white polygon in the residual map.
The peak around 110\arcsec\ corresponds to the far southwest 
excess labeled with a green polygon in the residual map.

Most of the structures in the inner region occur on a scale similar to
that of the radio sources in the center.
In order to study the radio/X-ray interaction in more detail,
Figure~\ref{fig:residual_chandra_halo} shows radio contours from
the VLA 1.5 GHz C-array 
(white contours)
and the 1.5 GHz B-array radio images \citep[green contours,][]{GBF+04}, 
overlaid on 
the central part of the {\it Chandra} residual map.
In general, the diamond-shaped outline of the extended radio emission (the
radio ``mini-halo'') corresponds
to a region of reduced X-ray emission, and there are excess X-ray regions
NNE and SSW of the edge of the mini-halo where the radio surface brightness
is dropping rapidly.
The mini-halo is roughly confined within 60\arcsec.  The distance from the 
center to the NE or SW corners of the diamond is about 50\arcsec; while to 
the NW or SE corners is about 70\arcsec.  
Note that in the interpretation of \citet{GBF+04}, 
the two radio bars are distinct from the diffuse radio emission, and in 
fact after their subtraction the morphology of the mini-halo becomes
roughly circular \citep[see Figure~6 of][]{GBF+04}.
However, in the residual image in \citet{GBF+04},
we can still see some excess radio emission in the NW and SE corners.
As noted previously, there is a significant excess associated with the
radio core and the southern radio jet.  The southern radio bar is a
region of X-ray excess, 
but the northern bar shows an excess to the east and a deficit to the west in 
X-rays.
The interpretation of the interaction between thermal and radio plasma 
will be discussed in \S~\ref{sec:discuss_radio} in detail.

The one major X-ray residual in the inner cluster which is not on the scale
of the central radio emission is the extended excess to the SW.
It is possible that this has been produced by the action of the central
radio source.
However, the SW excess is between the cD galaxy and the subcluster 
southwest of the main cluster center.
It is possible that this excess is the result of this merger or a previous
merger from this direction.
There is no obvious sharp surface brightness discontinuity associated with
this excess, which means that it either is not a shock or cold front,
or that the merger is not occurring primarily in the plane of the sky.
We note that \citet{MGW96} 
show that the line-of-sight velocity difference between the main 
cluster and the subcluster is $\sim 
2600$~km~s$^{-1}$, which is greater than the typical sound speed within 
clusters.
If the excess is the result of a merger, this suggests that the 
merger is mainly along the line-of-sight.

\section{Spectral Analysis}
\label{sec:spectrum}

\subsection{Temperature Profile}
\label{sec:spectrum:t_profile}

The azimuthally averaged, projected temperature profile for Abell~2626 was 
extracted
using both {\it XMM-Newton} and {\it Chandra} data
(upper panel of Figure~\ref{fig:ta-r_profile}). 
The spectra were fitted using an absorbed thermal model
(WABS*MEKAL) with XSPEC Version 12.3.1.
For the {\it XMM-Newton} data, all the spectra of the three cameras were
fitted simultaneously.
The energy bands used are 0.6-10.0 and 0.4-7.0 keV for
{\it XMM-Newton} and {\it Chandra} data, respectively.
Using 0.6-7.0 keV and 0.5-10.0 keV bands for {\it XMM-Newton} and {\it 
Chandra} data gives essentially the same results.
The spectra were grouped to have a minimum of 25 counts per bin.
We fixed the absorption to be $4.2\times 10^{20}$~cm$^{-2}$ \citep{DL90} 
and the 
redshift to be 0.0573 \citep{ACO89}.
Freeing these values gives essentially the same results.
Except the outermost data point where the reduced $\chi^2$ value is $\sim 
1.6$, all the other reduced $\chi^2$ values are $\la 1$, which indicates 
that the fits are good.
To ensure that the spectra are not affected by energy-dependent
PSF effects, the 
annuli of the {\it XMM-Newton} data were chosen such that all of the 
widths are at least twice the PSF
half energy width
of the PN camera.

The general trend is that the projected temperature increases from a 
central value 
of $\sim 2.5$~keV within 9\arcsec\ ($\sim 10$~kpc) to a 
maximum of $\sim 3.5$~keV at $100\arcsec-200\arcsec$ 
($110-220$~kpc).
Such a radial increase is normally seen in a cluster cooling core.
The projected temperature then decreases to $\sim 1.4$~keV at 
$\sim540\arcsec$ ($\sim 600$~kpc).
Interestingly, there is a sharp drop in the projected temperature from 
$\sim 
2.9$~keV at $210\arcsec$ (230~kpc) to $\sim 2.0$~keV at $270\arcsec$ 
(296~kpc) in the
{\it XMM-Newton} data.
Outside of this steep drop, the projected temperature gradient flattens.
The radius of the drop corresponds to the distance 
from the cD galaxy to the outer edge of the 
large excess emission in the southwest direction.
If the large excess emission indicates a merger signature, the temperature
jump might indicate that gas is heated within that region.
Unfortunately, the jump cannot be confirmed with {\it Chandra} as 
the {\it Chandra} data does not extend far enough in 
all directions to reach the distance of the jump seen in the azimuthally 
averaged temperature profile.  It only extends far enough towards the 
northwest direction, where there are not enough counts to constrain the 
temperature.

In general, the {\it XMM-Newton} data give lower values for the projected 
temperature profile. 
There are at least three possible explanations for this difference.
First, it might be due to the larger PSF of {\it XMM-Newton}.
The softer photons from the brighter, cooler inner regions may be
scattered to outer regions, hence making the outer regions appear cooler.
This effect may be seen from the second bin outward, and should be more
serious in regions where the gradient of the surface brightness is greatest
\citep[i.e., near the center;][]{Mar02}.
This might explain the discrepancy in the second and third bins of our
{\it XMM-Newton} data.
The second possibility is that there might have been a steady and mild 
enhancement
in the hard X-ray background during the {\it Chandra} observation
which was not removed by the standard cleaning procedure; 
this could make the {\it Chandra} emission appear hotter, particularly
in outer, low surface brightness regions.
Such an effect has also been noted by \citet{Mar02}. 
We have tried to assess the importance of such a background
enhancement by including a component for the hard background,
modeled as a power law with an exponential cutoff \citep{Mar02, Mar+03}
in the fits of the {\it Chandra} spectra.
However, this component did not significantly improve the fits, and the
best-fit amplitude of the component was negligible.
A third possibility is that the continuous accumulation of the contaminant 
layer on the {\it Chandra} ACIS optical blocking filter
was underestimated.

To study the true temperature of the cluster, we have also deprojected the 
temperature profile by using the XSPEC PROJCT model.  For each annulus, 
the model sums up the appropriate fractions of emissions contributed from 
the outer annuli.  The spectra of all the annuli were fitted 
simultaneously to get the deprojected temperature profile.  The middle 
panel of Figure~\ref{fig:ta-r_profile} shows the azimuthally averaged, 
deprojected temperature profile for Abell~2626 using both {\it XMM-Newton} 
and {\it Chandra} data.
For the {\it XMM-Newton} data, all the spectra from the three cameras were
fitted simultaneously.
The {\it XMM-Newton} and {\it Chandra} data are consistent 
within the 90\% confidence uncertainties, but the uncertainties of the 
{\it Chandra} data are much larger.  
Since the {\it XMM-Newton} data extends further out
and has smaller error bars, 
we describe results obtained from the {\it XMM-Newton} in the following. 
We also adopt the {\it XMM-Newton} spectral fits (lower panel of 
Figure~\ref{fig:ta-r_profile}) for the deprojection and 
the mass profile analysis below (\S\S~\ref{sec:density}, \ref{sec:time}, 
\& \ref{sec:mass}).

The deprojected central temperature is slightly lower than the projected 
one because the latter is contaminated by the hotter emission along the 
line of sight.  The difference is small since the hottest region has a 
temperature only 1 keV higher than the central one.  A more noticeable 
difference is that the temperature jump at around 240\arcsec\ is more 
obvious in the deprojected profile.  It drops from a temperature of $\sim 
3.5$ keV to $\sim 2.2$ keV.  The peak of the deprojected temperature 
profile is located just within the temperature jump ($\sim 180-240 
\arcsec$).

\subsubsection{Effects of XMM Background Uncertainties}
\label{sec:spectrum:background}

Since the cluster emission covers the whole field of view and 
the correct background cannot be estimated easily, the so called ``double 
background subtraction" method, common in {\it XMM-Newton} data analysis 
\citep{Arn+02}, cannot be applied.
The natural background subtraction method is to perform a 
renormalized blank-sky background subtraction without the second step of 
extracting a blank-field-subtracted spectrum in an emission free region 
in the ``double background subtraction" method. 
We use this ``single background subtraction" method.
Note that this method has also 
been applied to the study of a cluster with a similar redshift to Abell~2626 
\citep{ANA+01}.
As mentioned in \citet{ANA+01}, our treatment does
properly account for the cosmic-ray (CR) induced background.
However, the contributions from variations in
the local soft Galactic background (which depends on 
the position on the sky) and the extragalactic X-ray background component 
(which depends on the absorbing hydrogen column density along the line of 
sight and cosmic variance) were not included.  
In our case, the emission from Abell 2626 usually dominates the local 
soft Galactic background, and the CR induced background usually dominates 
the hard X-ray extragalactic background component.
Using the ROSAT All Sky Survey \citep{Sno+97},
we have compared the soft X-ray emission (0.47--1.21~keV) 
around Abell~2626, and the exposure-weighted emission in the regions used 
to extract the
blank-sky background, and the difference is less than 4\%.
The hydrogen column density towards Abell~2626 is about a factor of two 
higher than that for most of the blank-sky fields \citep{DL90}, but is not 
high enough to affect the cosmic X-ray background strongly.
Also, as mentioned previously, the CR induced background usually dominates 
the hard X-ray extragalactic background component.
Hence, the blank-sky backgrounds should represent the soft Galactic 
background and the hard X-ray background reasonably.  

We were concerned that, while the background renormalizations for MOS1 and MOS2
are modest ($\lesssim 7\%$ correction),
the renormalization of the PN background is larger ($\sim 26\%$ correction).
While the uncertainties in the background will affect most strongly
the derived temperatures of the outer regions of the cluster,
we emphasize here that most of our focus in this paper is on the X-ray and 
radio interaction in the innermost region where the results are 
nearly independent of background assumed.
We performed a number of tests to address the effects of the {\it 
XMM-Newton} background uncertainties on the deprojected temperature 
profiles.
We did this by:
1) varying the blank-sky background 
normalizations by $\pm$10\% ($\pm$20\%) for the MOS (PN);
2) using the blank-sky backgrounds without renormalization;
3) using the MOS1+MOS2 spectra only;
4) using the PN spectra only;
5) adding an extra 0.2~keV MEKAL component
(with normalization free and allowed to be negative in each annulus)
as a model for any variation in the soft Galactic component.
The test results are 
summarized in Figure~\ref{fig:ta-r_profile_bg}.
Our adopted deprojected temperature profile in this paper is shown in 
black diamonds.

Within $r < 180\arcsec$, all the deprojected temperatures are in very good
agreement.
All these tests show a significant jump in temperature near $240 \arcsec$ and a 
flattening of temperature beyond this jump.
As expected, increasing the background normalization (blue diamonds)
gave lower deprojected temperatures for the outer regions ($r > 180\arcsec$),
while decreasing the background normalization (green diamonds) gave higher
temperature.
Using the blank-sky backgrounds without renormalization (red crosses) also
gave higher temperatures (green diamonds).
The variation in the temperatures using different normalizations
was at most $\sim 1$~keV in the outer region.

The deprojected temperature profiles obtained from the MOS1+MOS2+PN 
spectra (black diamonds),
the MOS1+MOS2 spectra (magenta triangles),
and the PN spectra alone (cyan triangles) all agree reasonably well.
Thus, the PN data appear to give fairly reliable spectral information,
even though the uncertainty in the PN background normalization may 
be larger.

Finally, varying the amount of soft Galactic background (orange crosses)
gave deprojected temperatures in good agreement with those using the
renormalized blank sky backgrounds (black diamonds) except for the outermost
annulus.
The large deviation of the outermost data point may indicate that there
is less of soft Galactic emission in the direction of Abell~2626 than in 
the blank-sky fields because the fitted normalization of the extra soft 
Galactic background component in the model is negative.
The soft Galactic component should be uniform on the scale of a 
cluster.
However, a reduction in the soft Galactic background similar to that
suggested by the last annulus was found to be completely inconsistent with
the fits in the inner annuli.
Hence, we concluded that the reduced soft component in the 
outermost annulus was not due to a reduction in the soft Galactic background
towards the cluster.
The reduced soft component in the outermost annulus
may be the result of additional thermal structure in this region.

To summarize, all our results are independent of background used 
within $\sim 180 \arcsec$.
For the outer regions, the temperature profile is only slightly affected by
the choice of the background,
but the differences are mostly well within the error bars.
All the features described in the paper related 
to temperature measurements (e.g., the temperature jump, pressure, entropy 
profiles, etc.) are not affected.

\subsection{Central X-ray Spectrum and Cooling Rate}
\label{sec:cool}

We determined the radiative cooling rate by fitting the X-ray spectrum
of the central regions of the cluster from both {\it XMM-Newton} and {\it 
Chandra} data.
We extracted the spectrum from a circular region with a radius
of 65\arcsec,  which is roughly the cooling radius determined 
in \S~\ref{sec:time} below.
In XSPEC, we used the WABS*(MKCFLOW+MEKAL) model, which combines a cooling
flow component (MKCFLOW) with an isothermal model (MEKAL) to account for 
the gas in the outer regions.  
The absorption was fixed at the Galactic column of
$4.2 \times 10^{20}$~cm$^{-2}$ as before.
The upper temperature ($kT_{\rm high}$) and abundance of the MKCFLOW component
was set to be the same as the MEKAL component, which is expected if the gas
cooled from the ambient ICM.
Initially,  the lower temperature of the MKCFLOW component is set to be the 
lower limit for the model ($kT_{\rm low}=0.08$~keV), as expected for the 
classical cooling flow model in which the gas cools radiatively to very low
temperatures.
With this assumption, the best-fit cooling mass deposition rate is
$2\pm1 \, (3\pm3) \, M_{\odot}$~yr$^{-1}$ for {\it XMM-Newton}
({\it Chandra}) data in the 
0.6--10.0 (0.5--7.0)~keV energy band with a $\chi^2/dof=1.18$ (0.90).
The fitted upper temperature and abundance were
$kT_{\rm high} = 2.8\pm0.05 \, (3.3\pm0.2)$~keV
and $0.52\pm0.03 \, (0.9\pm0.1)$ solar, respectively.
Allowing $kT_{\rm low}$ to vary increased the mass 
deposition rate to be  $74\pm6 \, M_{\odot}$~yr$^{-1}$ for the {\it 
XMM-Newton} data, with $kT_{\rm low} = 1.5\pm0.1$~keV, $kT_{\rm high} = 
4.2^{+0.2}_{-0.5}$~keV, and an abundance of $ 0.45^{+0.03}_{-0.02}  $ 
solar, with a 
$\chi^2/dof=1.13$.
The {\it Chandra} data gave very poor constraints on the mass deposition 
rate or lower temperature with a $\chi^2/dof=0.87$
if $kT_{\rm low}$ was allowed to vary.
From the f-test, the probability that the fit to the model  with $kT_{\rm 
low}$ fixed is better than the fit with freeing $kT_{\rm low}$ is only 
$2 \times 10^{-11} (9 \times 10^{-3})$
for the {\it XMM-Newton} ({\it Chandra}) data.
For comparison, the mass deposition rate determined from {\it Einstein} 
data by surface brightness deprojection was $\sim 53 \, M_{\odot}$~yr$^{-1}$ 
\citep{WJF97}.
As has been found in many other clusters
\citep[e.g.,][]{PKP+03},
the spectral data for 
Abell~2626 can be fit with significant amounts of gas cooling, but only
by a factor of $\sim$3 in temperature.

\subsection{Density, Pressure and Entropy Profiles}
\label{sec:density}

Assuming the cluster is spherically symmetric, which appears to be roughly
true for Abell~2626, the radial surface brightness profile 
(Figure~\ref{fig:surface_brightness_xmm}) can be deprojected to give the
X-ray emissivity $\epsilon (r)$  and electron density $n_e (r)$ as
a function of radius
(Appendix~\ref{app1}).
The electron density profiles determined from {\it XMM-Newton} and {\it 
Chandra} data
are shown in the upper panel of Figure~\ref{fig:dps-r}.
Except for more fluctuations in the {\it Chandra} data, the two results
for radii greater than $\sim 7 \arcsec$ are in good agreement.
To avoid artifacts due to the assumption of zero emissivity outside the 
boundary during the deprojection \citep{TSB+03}, 
we have removed the 2 outermost points in each of the profiles plotted in 
Figure~\ref{fig:dps-r}.
In the {\it Chandra} data, we also note that the
two innermost data points are higher than the {\it XMM-Newton}
results by a factor of $\sim 1.5-2$.
Such a deviation in the innermost data points can also be seen when 
comparing the surface brightness profile of {\it Chandra} data to 
{\it XMM-Newton} data (not shown in this paper). 
The surface brightness profile, as well as other profiles, is 
centered at the location of the southwest nucleus of the cD galaxy. 
The central deviation between {\it Chandra} and {\it XMM-Newton} might be
due to the X-ray contribution of a 
central AGN (\S~\ref{sec:AGN}) and the differences in the PSFs of {\it 
XMM-Newton} and {\it Chandra}.
The smaller PSF of {\it Chandra} allows 
the central excess to be resolved more readily.
We did not remove the central source
during the analysis, since it is probably extended in the 0.3--10~keV band 
(see \S~\ref{sec:AGN}).
Also, including or excluding this region does not change any 
other conclusions in the paper.
The central excess
extends to $4\arcsec-7\arcsec$ in radius, which is larger than the PSF of 
{\it Chandra}.
This probably indicates that the excess is not due to an AGN alone.
The northeast nucleus of the cD galaxy is located 
at about $4\arcsec$ 
away from the central AGN (Figure~\ref{fig:2nuclei}). From the residual 
map 
(Figure~\ref{fig:residual_chandra_halo}), we 
clearly see that there is an extended excess around the northeast nucleus. 
There is also excess emission northwest of the AGN, which can also be 
seen in the soft band raw image of Figure~\ref{fig:2nuclei}.

The {\it XMM-Newton} data show that the electron density decreases from
$\sim 0.02$~cm$^{-3}$ at $\sim 7 \arcsec$ to $\sim 0.0004$~cm$^{-3}$ at 
$\sim 400 \arcsec$.
The density profile is generally smooth, with a small jump around 
$60\arcsec$ seen in the {\it Chandra} data.  However, the jump is not 
visible with the {\it XMM-Newton} data. 
This radius corresponds to the location where the logarithmic 
slope of the surface brightness profile changes, or equivalently,
to where the surface brightness goes from being dominated by the smaller to 
the larger of the two individual $\beta$ models.
We also noticed that the radio mini-halo is roughly confined within $\sim 
60\arcsec$.
Interestingly, this radius is also roughly coincident with the cooling radius 
(\S~\ref{sec:time} below).
Within the cooling radius and the region of the radio mini-halo,
radiative cooling and/or the effects of the relativistic particles and   
magnetic fields in the radio plasma may affect the density profile of   
the X-ray gas.

The gas density and temperatures profiles can be combined to give
the gas pressure $P= n k T$, where $n$ is the total number density in the 
gas (both ions and electrons).
We assume a fully ionized plasma with half cosmic abundances of heavy
elements which gives $n \approx 1.92 \, n_e$.
We also derive the gas entropy parameter $S \equiv k T/n_e^{2/3}$
\citep{PSF03}.
Since the spatial resolution of our density profile is much higher than the 
temperature profile, interpolation is used to determine the temperatures
on the finer density grid by fitting a high order polynomial (lower 
panel of Figure~\ref{fig:ta-r_profile}).
We assigned temperature errors 
of 5 (20)\% at radii  of smaller (larger) than $250\arcsec$.
Using a constant 
temperature model gives similar results, except for more negative
(unphysical) values for the total mass profile (\S~\ref{sec:mass} below).
Thanks to the nearly constant temperature profile of Abell~2626,
we believe that the interpolation of the temperature profile is
reasonably accurate.

The pressure profile (middle panel of Figure~\ref{fig:dps-r}) looks 
similar to the  
electron density profile, since the temperature is roughly constant. The 
feature at $60\arcsec$ becomes more
obvious, as the slope of temperature is negative in this region.
At around $240\arcsec$, there is a slope change in the pressure profile.  
This is associated to the temperature jump at that radius, which may be
a merger signature.

The entropy profile is shown in the lower panel 
of Figure~\ref{fig:dps-r}. 
There is a small bump in the entropy profile around 240\arcsec\ 
which is associated with the temperature jump and the change in slope of
the pressure profile in the same region.
The entropy may be boosted by a merger in that region.
Excluding this entropy bump region,
the logarithmic slope of the entropy profile in the outer regions
($\sim 100\arcsec-400\arcsec$) is 
$0.58\pm 0.06$,
which is significantly smaller
than the universal scaling relation slope of $\sim 1.1$
\citep[e.g.,][]{PSF03}. 
Note that there is a large scatter in the entropy profiles of
individual clusters
\citep[e.g., Figure~1,][]{PSF03},
and hence the smaller logarithmic 
slope might not be unusual.
There is a difference in the central entropy measured by {\it XMM-Newton} 
and {\it Chandra}.
This central entropy deviation comes from the difference in
central gas density measurements, which is probably due to the 
combination of a central excess in X-ray emission and differences in 
the PSFs of {\it Chandra} and {\it XMM-Newton}.
However, the spatial resolutions of the temperature measurements are not 
sufficient to determine if the entropy differences are due to the
temperature in the hot gas, or a central X-ray source.
In the following, we chose to present results of the {\it 
XMM-Newton} data with which the spatial resolution in density profile 
measurement is poorer.
The {\it XMM-Newton} data show that the entropy decreases from the outer    
value of $\sim 360$~keV~cm$^2$, eventually flattening to a value of $\sim
30$~keV~cm$^2$. If there is no feedback mechanism, a pure radiative 
cooling model will result in a power law entropy profile, with central 
entropy lower than $\sim 10$~keV~cm$^2$ within 10 kpc 
\citep[Figure~1 of][]{VD05}. 
The central entropy pedestal value of $\sim 30$~keV~cm$^2$ 
implies that there has to be heating in the central region, which can 
probably be explained by a central AGN.

\subsection{Cooling and Thermal Conduction Time Scales}
\label{sec:time}

Three important processes, heating, cooling and heat conduction,
determine the thermal structure in the hot gas
\citep[e.g.,][]{Sar86}.
While the
heating process in the central region is probably due to the AGN, the
other processes are related to the hot gas structure itself. Here we 
compare
two important time scales for the hot gas - the cooling and thermal
conduction time scales.
For the temperature gradients seen in typical cooling flow
clusters, conduction is likely to be an important process if it is
not suppressed by the magnetic field.

The cooling time was determined for each radius in the deprojected density
profile using the values of the deprojected density and the 
deprojected temperature for both {\it XMM-Newton} and {\it Chandra} data.
The rate of total emission from the gas was determined using the same
MEKAL model used to fit the spectra.
The integrated, isobaric cooling time $ t_{\rm cool}$ was determined.
A brief outline of the cooling time estimation procedure is given in 
Appendix~\ref{app2}.

We define the conduction time scale to be \citep{Sar86}
\begin{equation}
t_{\rm cond} \equiv \frac{n_e (T/|\nabla{T}|)^2 
k}{\kappa}
\, ,
\end{equation}
where 
$\kappa$ is the thermal conductivity
for a hydrogen plasma \citep{Spi62},
\begin{equation}
\kappa = 4.6\times10^{13} \left( \frac{T}{10^8 \, {\rm K}}
\right)^{5/2} \left( \frac{\ln \Lambda}{40} \right)^{-1} \, {\rm 
ergs~s}^{-1} \, {\rm cm}^{-1} \, {\rm K}^{-1} \, ,
\end{equation}
and $\Lambda=37.8+\ln[(T/10^8 \, {\rm K})(n_e/10^{-3} \, {\rm
cm}^{-3})^{-1/2}]$ is the Coulomb logarithm.

Figure~\ref{fig:time} shows the cooling and conduction time scales as a
function of angular radius.
The cooling time scale (plus symbols) decreases from $\sim 50$~Gyr at 
$\sim 
540\arcsec$, down to $\sim 1$~Gyr at $\lesssim 15\arcsec$.
Assuming the cluster has not changed much since $z \sim 1$, we define 
the cooling radius to be where $t_{\rm cool}$ is equal to the time since
$z = 1$, which is 7.7 Gyr.  
The cooling radius, $r_{\rm cool}$, is thus $\sim 60\arcsec-70\arcsec$ 
($\sim 66-77$~kpc).
Interestingly, $r_{\rm cool}$ is of the same size as the mini-halo 
(see \S~\ref{sec:residual}).
It is also located near the radius where 
the logarithmic slope of the surface brightness profile changes, or 
equivalently, where the surface brightness profile goes from being 
dominated by the smaller to the larger of the two individual $\beta$ 
models (\S~\ref{sec:SB}).  

The conduction timescale depends on the temperature gradient, which is
strongly affected by the uncertainties in the individual temperature 
points.
The error estimation for the conduction timescale is complicated by the 
fact that the adjoining deprojected temperatures used to derive the 
conduction timescale are correlated.
Hence, in calculating the uncertainties by error propagation,
we included the full covariance matrix for the temperatures obtained from the 
deprojection using XSPEC.
In the upper panel of Figure~\ref{fig:time}, the crosses
show the values of the conduction time derived using the individual temperature
values derived from annular {\it XMM-Newton} spectra 
(\S~\ref{sec:spectrum:t_profile}).
The conduction time derived from the {\it Chandra} data (lower panel of 
Figure~\ref{fig:time}) are in general shorter than those from the {\it 
XMM-Newton} data.  This may be due to the much larger scattering of the 
{\it Chandra} deprojected temperature profile.
The error bars of the {\it Chandra} data are much larger than that of the 
{\it XMM-Newton} data.
In order to reduce the effects of the uncertainties of the {\it Chandra} 
data, 
we also determine the conduction time scale in the inner region using the 
same interpolated temperature profile which was used to 
deproject the pressure/entropy profile
(\S~\ref{sec:density}).
The interpolated temperature profile was also used for the {\it 
XMM-Newton} 
data for comparison.
These values are shown as dots in Figure~\ref{fig:time};
the upper panel gives the values for the {\it XMM-Newton} data, and
the lower panel is for the {\it Chandra} data.
The {\it Chandra} conduction time scales are longer than the {\it
XMM-Newton} values in the innermost regions due to the higher densities
derived there from the {\it Chandra} data.

Here we describe results obtained from {\it XMM-Newton} since the error 
bars are much smaller than those from {\it Chandra}.
In the central regions of Abell~2626, the temperature gradient is very 
flat.
The overall temperature variation is only about 0.4 keV within the 
cooling radius of $\sim$65\arcsec. 
Both the best-estimate and the lower bound on the conduction time are 
larger than the cooling time in the central 
regions of the cluster.
The nearly isothermal temperature profile in Abell~2626 and resulting
low conduction imply that thermal conduction is not important compared
to cooling anywhere the cooling rate is high.
Hence, in Abell~2626 the thermal structure within the 
cooling radius of $\sim$65\arcsec\ 
is probably determined by cooling and heating processes only.
At a radius of $\sim$240\arcsec, the conduction time scale drops to $\sim 
0.3$~Gyr only, which is significantly shorter than the look-back time 
to a redshift $z \sim 1$.
Even including the uncertainty in the conduction timescale,
the upper limit on the time scale is still quite short.
This 
indicates that either the thermal structure (perhaps associated with a 
merger) happened within $\sim 0.3$~Gyr, or thermal conduction is 
suppressed at least by a factor of $\sim$20.
Except the feature at $\sim$240\arcsec, outside of $\sim$65\arcsec,  we 
cannot determine whether thermal 
conduction is important or not due to the large uncertainties.

\subsection{Mass Profiles}
\label{sec:mass}

Figure~\ref{fig:mass} shows the gas mass and total mass profiles of 
Abell~2626.
The gas mass profile is given by  
\begin{equation}
M_{\rm gas}(< r)
=
\int_0^r 4 \pi (r^\prime )^2 d r^\prime \rho_{\rm gas} ( r^\prime )
\, ,
\label{eq:mgas}
\end{equation}
where 
$\rho_{\rm gas} = \mu_e n_e$  is the gas mass density, and $\mu_e$ is
the mass per electron determined from the gas abundances.

The total mass of the cluster can also be determined if we assume
the hot gas is in hydrostatic equilibrium, and if the temperature of the
hot gas is known.
The total mass profile is given by \citep[e.g.,][]{Sar86}:
\begin{equation}
M_{\rm tot}(< r)
=
-\frac{kTr}{\mu m_p G}
\left( \frac{d \ln \rho_{\rm gas}}{d \ln r} + \frac{d \ln T}{d \ln r}  \right)
\, ,
\label{eq:hse}
\end{equation}
where $\mu$ is the mean mass per particle, and $m_p$ is the proton mass.
The temperature profile was interpolated from the {\it XMM-Newton} data
(Figure~\ref{fig:ta-r_profile}, \S~\ref{sec:density}).
Since the mass within any given radius is determined by local gas
properties and their derivatives,
the total mass can decrease or can even be negative.
(This occurred at $r = 57 \arcsec$ for {\it XMM-Newton} and at a number of 
points for {\it Chandra}, and these points were omitted from the plot.)
This can be due to measurement uncertainties in the data, which
are compounded by the derivatives, or by the breakdown of simple
hydrostatic equilibrium.
The radio mini-halo is within
$\sim60\arcsec$, and may be disturbing the hot gas or contributing
additional pressure support in this region (\S\S~\ref{sec:image_morph_cD}
\& \ref{sec:residual}).
Thus, the mass profile at small radii may not be represented accurately 
by hydrostatic equilibrium.
At radii of $r \ga 60\arcsec$, the general trend is for $M(<r)$ to 
increase with radius.

Compared to \citet{WJF97} rescaled to our value of the Hubble constant,
the gas masses we have obtained agree to within 2\% and 7\% 
at 321\arcsec\ and  489\arcsec, respectively.  The total mass agrees 
within 26\% at 321\arcsec.
The difference in the total mass at this radius is large, but the values
still agree within the uncertainties.
The two values of the total mass do not agree at 489\arcsec, probably due to
the larger uncertainty in the temperature in the outermost annulus.
The gas mass fraction
($M_{\rm gas}/M_{\rm tot}$) increases from $\sim 5 \%$ at $100\arcsec$ to 
$\sim 10 \%$ at $500\arcsec$.
The global trend is the same as is typical in relaxed clusters
\citep{DJF95,EF99,ASF02}.

\subsection{Hardness Ratio Maps}
\label{sec:spectrum:substructures}

To understand the rough spectral properties of different regions, we have 
created hardness ratio maps from both {\it XMM-Newton} and {\it Chandra} 
data
(Figures~\ref{fig:HRmap_xmm_large}, \ref{fig:HRmap_xmm_small}, 
\& \ref{fig:HRmap_chandra_halo}).
For {\it XMM-Newton} data, the PN image was not included because of its 
different spectral response and because the chip gaps and bad columns 
produced cosmetic artifacts in the smoothed image.
The hardness ratio maps were created by dividing the image in the hard band
(2--10~keV) by the image in the soft band (0.3--2~keV).
Each image was background-subtracted, exposure-corrected, and smoothed 
with the same scales as the $3 \sigma$ adaptively smoothed image of the
0.3--10~keV band. 
We also tried smoothing all images with the same scales as the $5 \sigma$ 
adaptively smoothed image of the 0.3--10~keV band, or with the same scales 
as the $3 \sigma$ adaptively smoothed image of the hard band (2--10~keV), 
and almost all the features look the same as the hardness ratio maps given
in this paper.
While some features are not consistently seen in both the {\it XMM-Newton} 
and {\it Chandra} hardness ratio maps, indicating the possibility that
they are statistic fluctuations in at least one of these maps,
the common features which show up in both maps in the inner regions
should be real
(Figure~\ref{fig:HRmap_xmm_small}).
The colors from black to blue to red to yellow to white represent the degree
of hardness from soft to hard.
Though the hardness ratio maps from {\it XMM-Newton} and {\it Chandra}
cannot be directly compared due to different 
spectral responses, the relative hardness is still useful to identify 
interesting regions.

\subsubsection{Global Structure}
\label{sec:spectrum:global}

Figure~\ref{fig:HRmap_xmm_large} is the hardness ratio map of the 
whole Abell~2626 from {\it XMM-Newton} data, overlaid with green contours 
of excess emission from the {\it XMM-Newton} residual map
(left panel of Figure~\ref{fig:residual_xmm}).
Part of the apparent inhomogeneity in hardness is just due to noise.
Also, near the blue and dark boundary of the hardness ratio map, 
the features start to be dominated by the hard background.
Although difficult to see due to the rather flat temperature profile, the 
overall picture is a soft core at the center.
At larger radii, the hardness 
increases outward to a radius indicated around the central cyan (light 
blue) circle, and then the hardness decreases outward. 
The central cyan circle is at a radius of 240\arcsec\ where the jump in
temperature was identified~(\S~\ref{sec:spectrum:t_profile}).
From the hardness ratio map, it seems that the 
harder gas is confined within the cyan circle.

The white circles in Figure~\ref{fig:HRmap_xmm_large} are the two extended 
X-ray emission regions associated with the merging subclusters
(\S~\ref{sec:image_morph_outer}). 
The discontinuities in the green contours near the white circles are
actually artifacts of chip gaps.
Both of the extended X-ray emission regions appear to be harder than their
surroundings, and each of them contains a softer point-like source. The 
soft source in the northeast extended X-ray emission region is located at 
the peak of the 
extended emission, and is probably associated with the galaxy 2MASX 
J23365372+2113322 (Table~\ref{table:source}), while the soft source in 
the southwest region is not associated with any identified galaxy.
The two soft point-like sources in the two extended emission regions
disappeared when the images were smoothed with the same scales as the $5 \sigma$
adaptively smoothed image of the 0.3--10~keV band, or with the same scales
as the $3 \sigma$ adaptively smoothed image of the hard band (2--10~keV).

\subsubsection{Central Region}
\label{sec:spectrum:central}

The details of the central region of the hardness ratio 
maps are shown in Figure~\ref{fig:HRmap_xmm_small}.
Contours of excess emission are shown
from the {\it Chandra} residual map 
(right panel of Figure~\ref{fig:residual_chandra}). 
While the {\it XMM-Newton} and {\it Chandra}
hardness ratio maps do not agree completely, probably due in part
to noise and in part to the larger
PSF of {\it XMM-Newton}, we can still identify some common features.

One interesting feature is that the S0 galaxy IC~5337 shows a softer tail
in the western direction.
On the {\it Chandra} hardness ratio map 
(right panel of Figure~\ref{fig:HRmap_chan_small}), 
there is a narrow tail of soft emission starting from the southern corner
of the IC~5337 tail and extending to the SWW.
We also see a softer region with a similar scale in the {\it 
XMM-Newton} hardness ratio map (left 
panel of Figure~\ref{fig:HRmap_xmm_small}).  

The northern excess region $\sim 30\arcsec$ above the cD galaxy (contours) 
appears softer, with 
an extension towards the northeast direction which can also be seen in 
Figure~\ref{fig:HRmap_xmm_large}.
If the region is near hydrostatic 
equilibrium and the material is supported by ideal gas pressure, we would 
expect the brighter (and hence denser) regions to have softer spectra.
This is roughly the case for the northern excess region. 
However, for the southern excess at a similar radius where the tongue is 
located (\S~\ref{sec:image_morph_cD}), the brighter 
excess region does not correspond to a softer region, 
which suggests a breakdown of hydrostatic equilibrium.
In particular, the X-ray excess tongue in the south direction of the cD 
galaxy (compared to Figure~\ref{fig:residual_chandra_halo}) appears to 
overlap both 
hard and soft regions extending in the same direction, making it hard to
understand its origin.

Between the main cluster and the southwest subcluster,
the large extended region of excess emission near the southwest corner
appears to have a harder spectrum in the {\it Chandra} data 
(right panel of Figure~\ref{fig:HRmap_chan_small}), 
while it appears to be soft in the {\it XMM-Newton} data
(left panel of Figure~\ref{fig:HRmap_xmm_small}).
From the hardness ratio map, it is not clear whether the extended 
excess is a shock heated region or a cold front.

In Figure~\ref{fig:HRmap_chandra_halo}, we can see clearly that there are 
two hard point sources located at the 
position of the cD galaxy IC~5338 and the S0 galaxy IC5337 (arrows).  The 
interpretation of two AGNs will be discussed in \S~\ref{sec:AGN}.
Radio contours \citep{GBF+04} are overlaid on the {\it Chandra}
hardness ratio map
to show in detail the complicated thermal and radio structures in the 
central region.
The southern radio bar appears to be in a harder 
region, while the northern one appears softer. 
If the radio bars are indeed thin tubes along the plane of the sky, 
the X-ray emission along the line-of-sights may mainly be contributed by 
the cluster emission beyond the central region.  In this case, the 
projected image or hardness ratio would not necessarily reflect the 
X-ray emissivities or hardnesses of the radio bars, and hence the lack of 
symmetry or a clear correspondence with features in the X-ray images or 
hardness maps might be understandable.

\subsection{Possible X-ray AGNs in IC~5338 and IC~5337}
\label{sec:AGN}

We searched for evidence of a central point source associated
with  the cD galaxy IC~5338, which may indicate the existence of an X-ray AGN.
When {\it wavedetect} was run on the 0.3--10~keV and 0.3--2~keV band {\it 
Chandra} images (\S~\ref{sec:sources_detection}), each detected one 
central source located at the southwest and northeast nucleus of the cD 
galaxy, respectively. 
Both were possibly extended.
We also ran {\it wavedetect} on the hard band (2--10~keV) image,
but it did not detect a central point source.
However, when we inspected the hard band image, we noticed a possible
hard point source.
This source can be seen in the {\it Chandra} hardness ratio map
(Figure~\ref{fig:HRmap_chandra_halo}).
We extracted the {\it HST} image of the center of IC~5338 from the
archive.
The 300 s exposure was taken on 1999 June 26 using the {\it HST} WFPC2 camera 
with the F555W filter.
Figure~\ref{fig:2nuclei} shows clearly that this cD galaxy has 
two distinct nuclei, with the northeast nucleus being brighter optically.
The two optical nuclei are located at the peaks of the soft X-ray image,
with the radio core source in IC~5338 centered on the hard X-ray, 
southwest nucleus.
The hard X-ray point source is only separated from the peak of the radio
core by less than 1\arcsec.
Thus, we believe that this source is the AGN, and that the 
point-like hard X-ray source arises from this AGN.

To study the X-ray properties of the IC~5338 AGN,
{\it Chandra} data were used to do photometry.
Due to the limited number of photons, we cannot fit the spectrum
of the AGN.
Hence, we estimated the properties of the cD AGN by assuming that it has a 
power law spectrum with a photon index $\Gamma$.
We extracted the total photon counts in a circular region with a 3 pixel
radius in both
hard (2--10~keV)
and soft (0.3--2~keV) bands.
In order to subtract the cluster emission as well as true background,
we took background from a 3-pixel-wide annulus just outside the central region.
The program {\it PIMMS}\footnote{http://heasarc.gsfc.nasa.gov/Tools/w3pimms.html}
was used to determine
the photon index and the unabsorbed flux which gave the correct 
hardness ratio and photon counts.
We assumed that absorbing column was given by the Galactic value
$N_H = 4.2 \times 10^{20}$ cm$^{-2}$ \citep{DL90}.
The best fit photon index is
$\Gamma = 2.2^{+0.7}_{-0.4}$.
The unabsorbed flux in the 0.3-10.0~keV band is
$F_X = 1.6^{+0.7}_{-0.4} \times10^{-14}$ ergs cm$^{-2}$ s$^{-1}$, 
which gives an X-ray luminosity of
$L_X = 1.2^{+0.5}_{-0.3} \times10^{41}$~ergs~s$^{-1}$ in the same
energy band.

We also searched for evidence for a point source associated with an
AGN in the center of the X-ray bright S0 galaxy IC~5337.
In this case, {\it wavedetect} did detect a point source in the
{\it Chandra} hard band (2--10~keV) image at the center of IC~5337.
This point-like, hard source is clearly seen in the 
hardness ratio map (Figure~\ref{fig:HRmap_chandra_halo}).
The S0 galaxy is also a radio source, and the X-ray point source is
apparently associated with this central AGN.
Again, the point source was too faint for spectral fitting.
Using the same method described above,
we determined the best-fit spectral index and flux
assuming the Galactic hydrogen column density of
$4.2\times10^{20}$~cm$^{-2}$.
The best fit photon index is  $\Gamma = 0.2^{+0.6}_{-0.3}$, the unabsorbed
flux in the 0.3--10.0~keV band is
$F_X = 3.2^{+1.9}_{-1.7} \times10^{-14}$ ergs cm$^{-2}$ s$^{-1}$,
and the X-ray luminosity is
$L_X = 2.4^{+1.5}_{-1.3}\times 10^{41}$~ergs~s$^{-1}$.
This spectrum index seems to be low (hard) for a typical AGN.
One possibility is that there is excess internal absorption associated
with the S0 galaxy AGN.
To assess this,
we fixed the photon index to $\Gamma = 1.5$.
In this case, the best-fit hydrogen column density is
$N_H = 1.3^{+0.7}_{-0.8} \times 10^{22}$~cm$^{-2}$, the unabsorbed flux 
is 
$F_X =  ( 3.0 \pm 1.5 ) \times 10^{-14}$ ergs cm$^{-2}$ s$^{-1}$, and 
the X-ray luminosity is
$L_X =  ( 2.3 \pm 1.1 ) \times 10^{41}$~ergs~s$^{-1}$.
Somewhat counterintuitively, 
the extra absorption gives a lower unabsorbed flux 
because of the steeper spectral index assumed.

\section{Kinematics of the S0 Galaxy IC~5337}
\label{sec:discuss_S0}

The bow-shock-like shape 
suggests that the S0 galaxy is moving east directly towards the cluster 
center (right panel of Figure~\ref{fig:S0}).
The hardness ratio maps of the center of the cluster
(Figures~\ref{fig:HRmap_xmm_small} \& \ref{fig:HRmap_chandra_halo})
show a region of soft (and presumably, cool) X-ray emission extending to
the west of the S0 galaxy.
This geometry would be consistent with the S0 galaxy falling into the
cluster from the west;
the cool tail of gas would be either interstellar or intragroup gas
associated with this galaxy which is being stripped by ram pressure from
the motion through the ICM of Abell~2626.
The steepness of the intensity gradient in the cluster radio mini-halo 
on the western side compared to the eastern side 
(left panel of Figure~\ref{fig:smoothed_imageChandra}) may also indicate that 
it is
compressed by the S0 galaxy.

The radio structure complicates the interpretation of the motion of the
galaxy, since there is a component centered on the S0 galaxy, and a 
component located at a position of about 190\arcdeg\ (measured from 
north to east) of the S0 galaxy on the 1.5 GHz B-array map
(Figures~\ref{fig:smoothed_imageChandra}, 
\ref{fig:residual_chandra_halo}, \& \ref{fig:HRmap_chandra_halo}).
The 1.5 GHz C-array map shows three radio tails in the south, southwest 
and west directions.
If the south component is a Narrow-Angle-Tail (NAT) behind the S0 galaxy,
then it is likely to be moving to the NNE (about 10\arcdeg\ 
measured from north to east) direction.
On the other hand, the other two components (west and southwest tails) 
are more consistent with the bow-shocked structure in the X-ray image 
which indicates that the S0 galaxy is moving to the east.

The interpretation of the kinematics of IC~5337 is confused by an
ambiguity in the literature concerning its radial velocity.
There are two optical determinations of the radial velocity which give
values which are consistent with the velocity of other galaxies in
the center of Abell~2626, including the cD galaxy IC~5338.
\citet{HVG99} give a velocity of 16,485$\pm37$ km s$^{-1}$, and
\citet{Fal+99} find 16,562$\pm60$ km s$^{-1}$.
These values are consistent with the 21 cm line determination of
16,485$\pm40$ km s$^{-1}$ by \citet{GP93}.
On the other hand,
\citet{MGW96} list an optical velocity of 18,903$\pm39$ km s$^{-1}$, and 
\citet{KK81} 
list a velocity of 19,097 km s$^{-1}$ with an uncertainty of the 
order of 100 km s$^{-1}$.
The difference between these velocities and the others is much greater
than the quoted (or any likely) measurement errors.
The velocities given by \citet{MGW96} and \citet{KK81} would be inconsistent
with IC~5337 being a part of the main central cluster Abell~2626.
On the other hand, these higher velocities are consistent with
IC~5337 being a member of the southern subcluster with a
mean velocity of 19,164$\pm138$ km s$^{-1}$
\citep{MGW96}.
Most of the other members of this subcluster lie considerably
to the south. 

If IC~5337 is a member of the southern subcluster, its dynamics 
would probably follow somewhat the motion of the subcluster. We would 
expect IC~5337 would be infalling from SSW to NNE (towards the 
position angle of 30\arcdeg\ direction measured from north to east). 
This would be inconsistent with our interpretation of the bow-shaped X-ray
region to the east of IC~5337.
Based on the X-ray image, we suggest that IC~5337 is falling directly
into the main cluster from the west,
and may not be associated with the southern subcluster.
On the other hand, given that there are two velocity measurements consistent
with the kinematics of the southwest subcluster \citep{MGW96, KK81}, it 
seems unlikely that
both are in error in the same way.
One possible explanation is that there are two galaxies 
located along the 
line-of-sight toward IC~5337.
Indeed, on the {\it Chandra} X-ray image, there 
are two point sources identified within the extended optical emission of 
IC~5337.
One is located at the center of the S0 galaxy with two west and 
southwest NATs, while 
the other is located at its southern edge with a south NAT.
The southern X-ray source was removed during the analysis 
related to the S0 galaxy in this paper.  
We have accessed the public available IR\footnote{2MASS: 
http://irsa.ipac.caltech.edu/Missions/2mass.html}, 
optical\footnote{The STScI Digitized Sky Survey: 
http://stdatu.stsci.edu/cgi-bin/dss\_form} and UV\footnote{GALEX: 
http://www.galex.caltech.edu/} 
surveys to search for a point source at the location of the southern 
X-ray point source.  
While there is no obvious emission in the IR H and K band images, there 
seems to be extended emission associated with the southern X-ray source in 
the IR J band, optical band and the UV band images.
However, one cannot easily determine if this emission is a separate galaxy
or a feature in the disk of the S0 galaxy.

\section{Cooling Flow Interaction with the Central Radio Source?}
\label{sec:discuss_radio}

\begin{figure}
\includegraphics[angle=0,width=8.2cm]{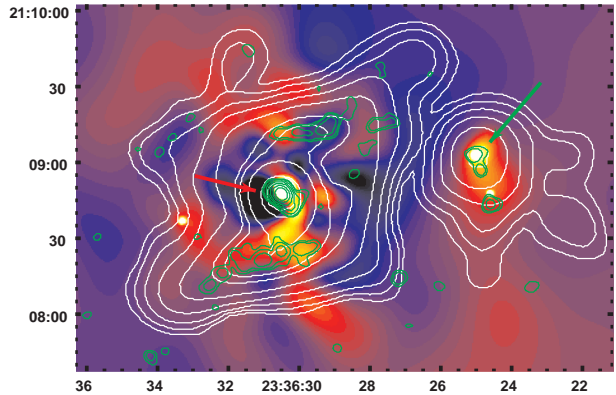}
\caption{Residual map from {\it Chandra} observation of the central  
$3\farcm5\times2\farcm4$ region
of Abell~2626,
generated from a background-subtracted image.
The color scale from yellow to red to blue to black represents
the residuals, running from excess to deficit.
The white solid lines are 1.5 GHz C-array radio contours showing the 
mini-halo.
The green solid lines are 1.5 GHz B-array radio contours showing the
radio bar structures.
Both radio contours were taken from \citet{GBF+04}.
The arrows on the left and right indicate the cD galaxy IC~5338 and the S0 
galaxy IC~5337, respectively.
}
\label{fig:residual_chandra_halo}
\end{figure}

\begin{figure}
\includegraphics[angle=270,width=8.2cm]{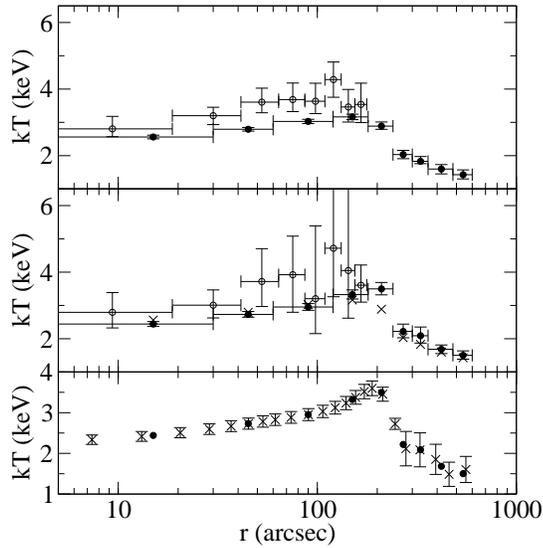}
\caption{
{\it Upper Panel}:
Profile of the azimuthally averaged, projected temperature $T$ of 
Abell~2626. The {\it XMM-Newton} ({\it Chandra}) data are shown 
as solid (open) circles.
{\it Middle Panel}:
Profile of the azimuthally averaged, deprojected temperature $T$ of 
Abell~2626. The {\it XMM-Newton} ({\it Chandra}) data are shown 
as solid (open) circles.
The black crosses are the projected temperatures from the {\it XMM-Newton}
data in the upper panel for comparison (without error bars for clarity).
In both the upper and middle panels, the error bars are at the 90\%
confidence level.
{\it Lower Panel}: Crosses are the temperatures calculated 
from interpolation 
of the original {\it XMM-Newton} deprojected temperature profile; this 
interpolated
profile was used for 
the pressure/entropy/time scales deprojection and the mass profile analysis 
in \S\S~\ref{sec:density}, \ref{sec:time}, \& \ref{sec:mass}.
The interpolated temperature profile for 
the {\it Chandra} data analysis is similar but not shown for clarity.  
The errors are assigned as 5 (20)\% at radii of smaller (larger) than 
250\arcsec\ to include possible systematic errors.  The solid circles are 
the original {\it XMM-Newton} measured deprojected temperatures as shown 
in the middle 
panel, without error bars for clarity.
}
\label{fig:ta-r_profile}
\end{figure}

\begin{figure}
\includegraphics[angle=270,width=8.2cm]{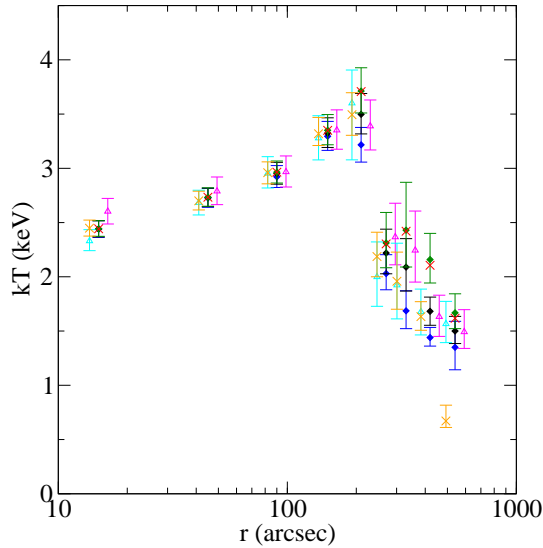}
\caption{
Profile of the azimuthally averaged, deprojected temperature $T$ of 
Abell~2626 with different {\it XMM-Newton} backgrounds.
The black diamonds give our adopted {\it XMM-Newton} temperature profile
(the black solid circles in the second panel of 
Figure~\ref{fig:ta-r_profile}).
The other temperature profiles are 
generated by:
1a) (blue diamonds) increasing the blank-sky background normalizations by 
10\% (20\%) for MOS (PN); 
1b) (green diamonds) decreasing the blank-sky background normalizations by 
10\% (20\%) for MOS (PN); 
2) (red crosses) using the blank-sky backgrounds without renormalization
(no error bars for clarity);
3) (magenta triangles) using the MOS1+MOS2 spectra only (data shifted to
the right for clarity);
4) (cyan triangles) using the PN spectra only (data shifted to the left
for clarity);
5) (orange crosses) adding an extra 0.2~keV MEKAL component to represent
variations in the soft Galactic component (normalization free and
allowed to be negative in 
each annulus, data shifted to the left for clarity).
The error bars are at the 90\% confidence level.
}
\label{fig:ta-r_profile_bg}
\end{figure}

\begin{figure}
\includegraphics[angle=0,width=8.2cm]{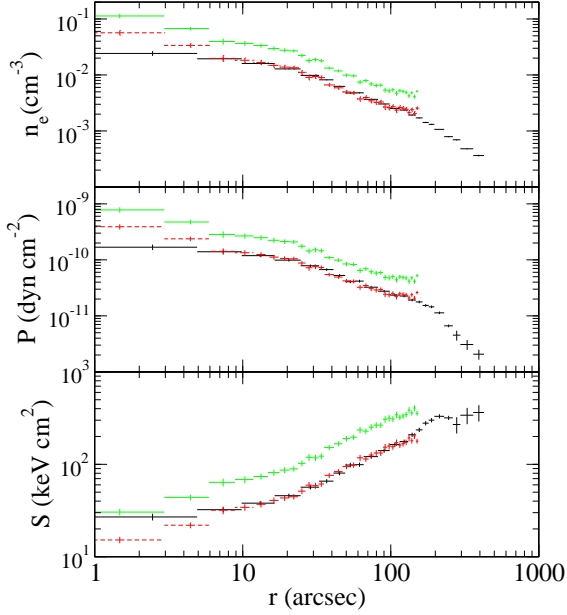}
\caption{{\it Upper Panel}: Azimuthally averaged electron density 
profile of Abell~2626.
The {\it XMM-Newton} ({\it Chandra}) data are shown 
as black solid (red dashed) symbols.
For clarity, we have also multiplied
the {\it Chandra} data and error bars
by a factor of 2 (green solid symbols).
The error bars are at the 1 $\sigma$ level.
{\it Middle Panel}: Azimuthally averaged pressure profile,
with the same notation as the upper panel.
{\it Lower Panel}: Azimuthally averaged entropy profile,
with the same notation as the upper panel.
}
\label{fig:dps-r}
\end{figure}

\begin{figure}
\includegraphics[angle=270,width=8.2cm]{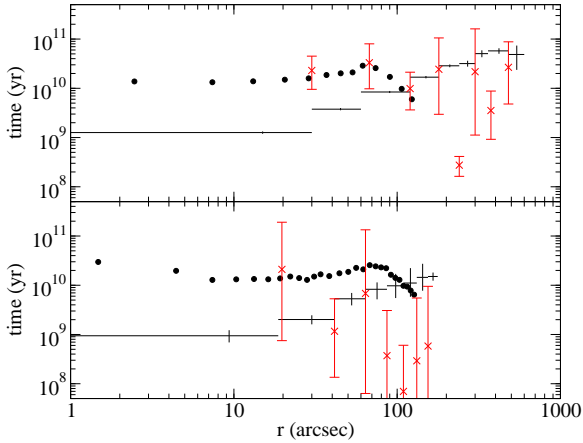}
\caption{{\it Upper Panel}: Cooling and thermal conduction time scales
in Abell~2626 obtained from the {\it XMM-Newton} data. 
The cooling times are the plus symbols with horizontal bars showing
the annular regions sampled.
The conduction time scales derived directly from individual data points are
the crosses with larger vertical error bars.
The dots are conduction time scales calculated by interpolating the
temperature profile of the cluster
(see text in \S~\ref{sec:time}).
The radial regions for the conduction time are not given for clarity.
The error bars are at the 1 $\sigma$ level.
{\it Lower Panel}: Same as the upper panel, but with the time scales
calculated from the {\it Chandra} data. 
}
\label{fig:time}
\end{figure}

\begin{figure}
\includegraphics[angle=270,width=8.2cm]{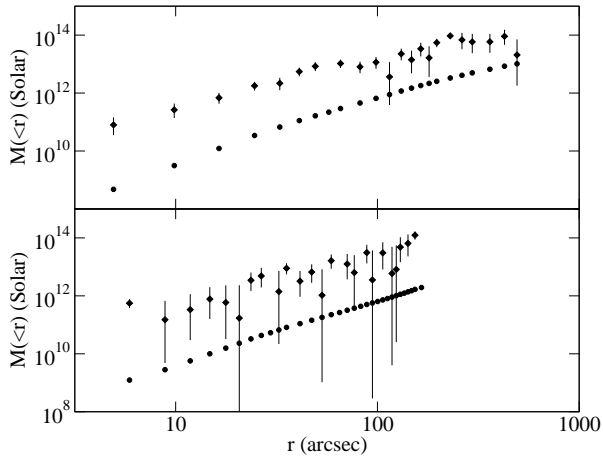}
\caption{{\it Upper panel} is the interior mass profile $[M(<r)]$,
of Abell~2626 obtained from {\it XMM-Newton} data, while the {\it lower 
panel}
is from {\it Chandra} data.
The total mass is shown with diamond symbols,
while the gas mass points are circles.
The error bars on the masses are calculated using propagation of errors 
at 1 $\sigma$ level.
The error bars on the gas mass are plotted, but are smaller than the circle
symbols.
}
\label{fig:mass}
\end{figure}

\begin{figure}
\includegraphics[angle=0,width=8.2cm]{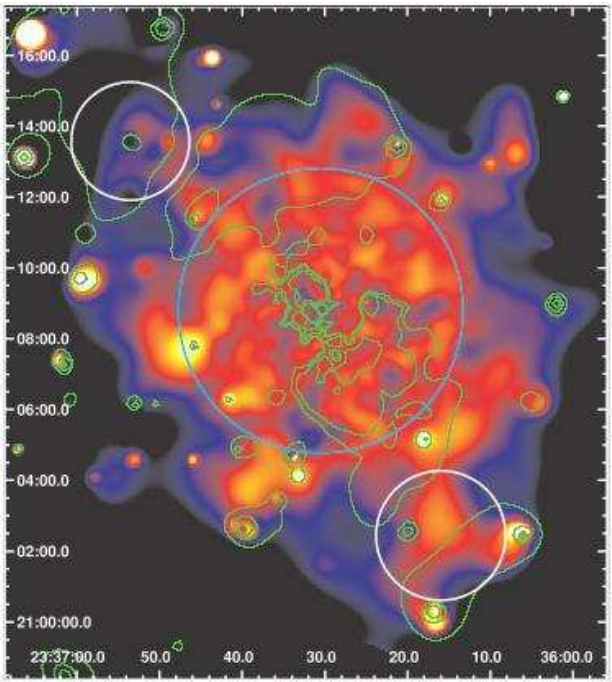}
\caption{
Hardness ratio ($HR$) map of Abell~2626 from the {\it XMM-Newton} 
observation.
Background was subtracted from the images combined to calculate the
hardness ratio.
The color scale ranges from black ($HR \sim 0$) to blue ($HR \sim 0.1$) to red 
($HR \sim 0.3$) to yellow ($HR \sim 0.5$) to white ($HR \sim 2.5$),
showing increasingly hard emission.
The green solid lines are contours from {\it XMM-Newton} residual 
map
(left panel of Figure~\ref{fig:residual_xmm}) showing positive residuals 
(excess 
emission).
The white circles are the regions of extended X-ray emissions to the
northeast and southwest.
The cyan circle has a radius of 240\arcsec\ corresponding to the jump in 
the temperature
profile.
}
\label{fig:HRmap_xmm_large}
\end{figure}

\begin{figure}
\includegraphics[angle=0,width=5in]{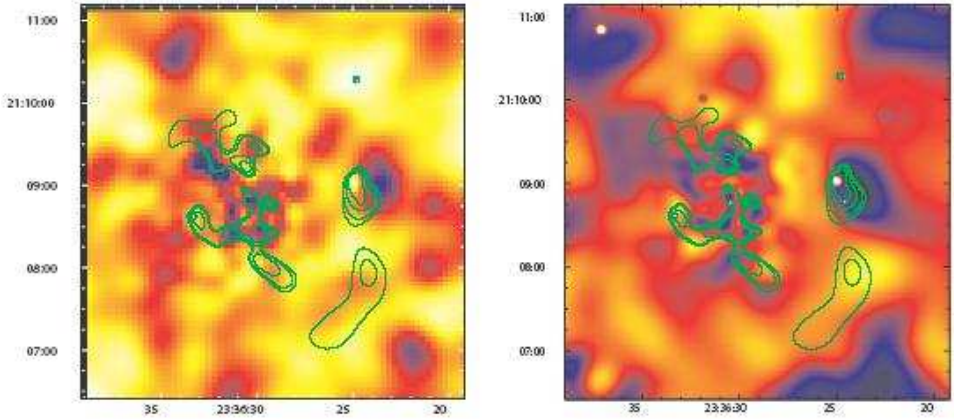}
\caption{
{\it Left panel:}
Central portion
($4\farcm6 \times 4\farcm7$) 
of the {\it XMM-Newton} hardness ratio map 
(Figure~\ref{fig:HRmap_xmm_large}).
The color scale is similar to that in  
Figure~\ref{fig:HRmap_xmm_large}, but with the hardness ratio extremes 
being $\sim 0.2$ and $0.35$ (dark blue to white yellow).
The green solid lines are contours from the {\it Chandra} residual map
(right panel of Figure~\ref{fig:residual_chandra})
showing the excess emission.
{\it Right panel:}
Hardness ratio map from the {\it Chandra} data, for the same regions and
with the same contours as 
the left panel.
Background was subtracted from the images combined to calculate the
hardness ratio.
The color scale is similar to that in  
Figure~\ref{fig:HRmap_xmm_large}, but with the hardness ratios in the
range from 
$\sim 0.1$ to 1.5 (dark blue to white yellow).
The cD galaxy IC~5338 and the S0 galaxy IC~5337 are located at (23:36:31, 
+21:08:47) and (23:36:25, +21:09:03), respectively.
\vskip 15mm
}
\label{fig:HRmap_xmm_small}
\label{fig:HRmap_chan_small}
\end{figure}

\begin{figure}
\includegraphics[angle=0,width=8.2cm]{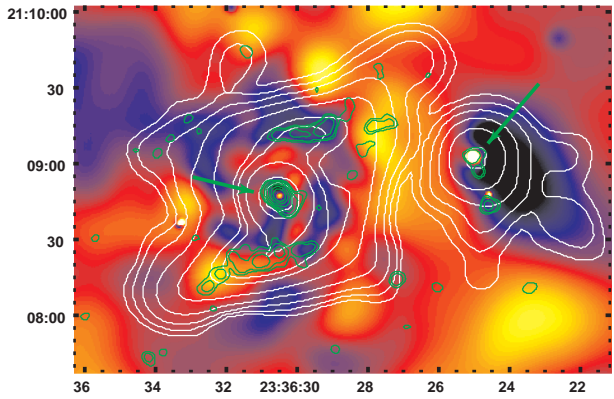}
\caption{Central region ($3\farcm5\times2\farcm4$) of the hardness ratio
map from the {\it Chandra} observation.
Background was subtracted from the images combined to calculate the
hardness ratio.
The color scale is roughly the same as in 
the right panel of Figure~\ref{fig:HRmap_chan_small}.
The white solid lines are 1.5 GHz C-array radio contours showing the 
mini-halo.
The green solid lines are 1.5 GHz B-array radio contours showing the
radio bar structures.
Both radio contours were taken from \citet{GBF+04}.
The arrows on the left and right indicate the cD galaxy IC~5338 and the S0 
galaxy IC~5337, respectively.
}
\label{fig:HRmap_chandra_halo}
\end{figure}

The most unusual features of the radio image of Abell~2626 are the two 
elongated
radio bars to the north and south of the radio core
(Figures~\ref{fig:smoothed_imageChandra}, 
\ref{fig:residual_chandra_halo}, \& \ref{fig:HRmap_chandra_halo}).
Their symmetric positions suggests they are radio lobes, but their
elongated shapes are unusual.
By comparison to radio lobes associated with other cooling core dominant
radio galaxies, one might expect these to be regions of reduced
X-ray emission surrounded by bright rims
\citep[``radio bubbles'',][]{Fab+00,BSM+01}.
In radio bubbles, the radio plasma has apparently displaced the X-ray gas.
One would expect the level of reduction in the X-ray surface brightness
to depend on
the extent of the bubbles along the line-of-sight.
If they are highly oblate flattened regions with a large extent along the
line-of-sight, the reduction in the X-ray surface brightness might be
quite significant.
If they are highly prolate tubes, then the reduction would be small.
In fact, there is no obvious correlation between the two radio bars
and the {\it Chandra} X-ray image, residual map or hardness ratio map in 
Figures~\ref{fig:smoothed_imageChandra},
\ref{fig:residual_chandra_halo}, \& \ref{fig:HRmap_chandra_halo}, 
respectively.
This may indicate that the radio bars are thin tubes, with a small
extent along the line-of-sight, or that the radio plasma is mixed with
the X-ray gas.
In the numerical simulations by \citet{Rob+04}, if the magnetic field of
the bubble is weak, the bubble is very unstable and can be destroyed easily,
leaving no obvious density contrast.
The ratio of gas to magnetic pressure in the center of Abell~2626 is of 
the order of 100 \citep{RLB+00, GBF+04}, which is the
same order as the weak
magnetic field in the simulation by \citet{Rob+04}.
However, it should be noted that the magnetic field in the radio bars could 
be much higher.
One concern with this model is that it seems unlikely that such mixing would
preserve the narrow radio bars which are observed.

One possible interpretation of the arc-like X-ray feature
near the center of Abell~2626
(upper right panel of Figure~\ref{fig:2nuclei})
is that the central cD galaxy is moving to the west relative to
the local intracluster gas.
This could explain the bending of the X-ray emission near the
center and the inner radio jets, as well as the elongation of the
radio bars.
However, in this case one might expect that the western edges of the
two radio bars would indicate the positions where the radio jets have 
stopped.
These positions are not mirror symmetric about the position of the radio 
core, as might be expected for jets perpendicular to the motion of the  
nucleus of the cD galaxy.
Even if the jets are not perpendicular to the motion of the nucleus, one 
of the western edges of the two radio bars should have been behind the 
nucleus of the cD galaxy in this interpretation.

There are two X-ray excess ``tongues'' from the central cD galaxy to the 
two radio bars, with the southern ``tongue'' appears to be stronger 
(upper right panel of Figure~\ref{fig:2nuclei}, \& 
Figure~\ref{fig:residual_chandra_halo}).
A similar tongue has been seen in Abell~133, where it was suggested that
it might arise from a cluster merger,  
Kelvin-Helmholtz instabilities around the core, the buoyant uplift from
a radio bubble, or a cooling wake \citep{FSR+04}.
Although subclusters may be merging with the outer regions of Abell 2626,
there is no clear evidence for merger distortion in the central region.
Even though the two cD nuclei may indicate a previous merger,
the two nuclei are well inside the cD galaxy, which means
that any associated merger may have happened a long time ago.
Also, a merger might be expected to produce a tongue and radio lobe 
on only one side of the nucleus, as in Abell 133.
For Abell~2626, the mean line-of-sight velocity of the galaxies is 
16,533 km s$^{-1}$, while that of the central cD galaxy is 16,562 km s$^{-1}$
\citep{MGW96}, so the radial velocity of the cD galaxy relative to the 
cluster
is low.
Moreover, there is no obvious cold or shock front, which means that any 
motion
in the plane of the sky is also slow.
Therefore, there is no 
obvious velocity for the cD galaxy to induce Kelvin-Helmholtz 
instabilities.
If the tongue is caused by a cooling wake, there would be only one.  But 
we see two of them, though the northern one is much weaker than the 
southern one.

Eliminating the explanations by a cluster merger,
Kelvin-Helmholtz instabilities 
and the cooling wake,
the tongue seems to be best explained by the buoyant uplift from a radio 
bubble. 
In fact, a 
recent simulation indicates that when a bubble rises in the intracluster 
medium, large rolls extend out and converge under the bubble. This results 
in an upwelling and compression of material under the wake of the 
bubble \citep{Rob+04}.
The morphology of the southern tongue shown in
Figure~\ref{fig:residual_chandra_halo} looks very similar 
to the simulated result \citep{Rob+04} if we assume that the radio bar 
is the head of the plume.
However, if the southern ``tongue'' is caused by uplifted cool gas,
why is the northern X-ray tongue associated with the northern radio 
bar much weaker, given the two radio bars are similar in appearance?
While there is no radio flux measurement to the radio bars in the 
literature, the stronger of the radio emission from the southern bar can 
be noted on the VLA 330 MHz B+DnC-array radio map of Figure~4 in 
\citet{GBF+04}.  This may explain the stronger uplift of the 
thermal plasma in the south.

Jet precession might also provide a natural explanation of the structure
of the central region.
The southwest cD nucleus is an AGN (see \S~\ref{sec:AGN}), and may be
ejecting two jets towards the north and south.
We assume the two jets are precessing about an axis which is nearly
perpendicular to the line-of-sight along the north-south direction, so
that the jets sweep out two conic surfaces to the north and south.
If the two jets are stopped at approximately equal radii from the
AGN (at a ``working surface''), and if radio emission is produced by
particles accelerated when the jets are stopped, then the radio bars
might be produced as the jets precessed.
If the jets are narrow, then the radio emission at the working surface may
be narrow, explaining the shape of the radio bars and also
the lack of anticorrelation between the two radio bars and 
the X-ray emission.
The precession of the jets might be due to the gravitational effects of
the second cD nucleus to the north, and this nucleus might have disrupted
or weakened the northern X-ray tongue.
Alternatively, the dominance of the southern X-ray tongue may simply be due 
to the southern jet being more powerful.
There is also evidence of jet precession in other cooling 
core clusters \citep{GFS06}.
Jet precession has also been suggested to be a requirement for the AGN 
feedback mechanism to solve the cooling flow problem \citep{VR06}.

Abell~2626 is one of the clusters associated with a possible radio mini-halo, 
an extended diffuse radio source in the cooling core region of the 
intracluster
medium \citep{GBF+04}.
In the cooling core region, the magnetic field strength 
is expected to be high due to compression. Consequently the radiative 
cooling time of the relativistic electrons is expected to be as short as 
$\sim 10^7-10^8$~yr, which is much shorter than the time needed to travel 
across the extended region \citep{GBF+04}. 
In order to explain the existence of the radio mini-halo, 
an {\it in-situ} re-acceleration mechanism is needed.
The steep radio spectrum of the mini-halo
\citep[$\alpha \sim 2.4$,][]{GBF+04} may be due to confinement by the
X-ray gas and radiative losses by the relativistic electrons.

While there is no very obvious correlation between the radio emission 
from the mini-halo and the X-ray emission from the X-ray 
images (Figure~\ref{fig:smoothed_imageChandra}), 
there is some evidence of interaction between the radio 
mini-halo and the surrounding intracluster medium in the residual maps 
(Figure~\ref{fig:residual_chandra_halo}).  
In general, the diamond-shaped outline of the extended radio emission (the
radio ``mini-halo'') corresponds
to a region of reduced X-ray emission, and there are excess X-ray regions
NNE and SSW of the edge of the mini-halo where the radio surface brightness
is dropping rapidly.
The mini-halo is roughly confined within 60\arcsec.  The distance from the 
center to the NE or SW corners of the diamond is about 50\arcsec; while to 
the NW or SE corners is about 70\arcsec.  
As noted previously (\S~\ref{sec:residual}),
in the interpretation of \citet{GBF+04}
the two radio bars are distinct from the diffuse radio emission, and in 
fact after their subtraction the morphology of the mini-halo become 
roughly circular \citep[see Figure~6 of][]{GBF+04}, though we can still 
see some excess radio emission in the NW and SE corners.
Interestingly, there is also a 
small density jump in the X-ray gas around 60\arcsec\ (\S~\ref{sec:density} above). 
The size of the mini-halo is also coincident with the cooling radius 
(\S~\ref{sec:time} above).
This supports the re-acceleration model for the origin of the 
mini-halo \citep{GBF+04}, though it could just be a coincidence since the
definition of the cooling radius is somewhat arbitrary.

An excess in the residual map corresponds to a region where the gas 
density 
is higher than in the symmetric model 
(Figure~\ref{fig:residual_chandra_halo}).
The shape of the mini-halo, if it is really 
elongated in shape (after proper subtraction of the radio bars),
may be due to variations in the X-ray gas density.
The radio source may be more strongly confined by the gas
in the NE and SW directions.
The points on the diamond-shaped radio mini-halo to the NW and SE
may indicate regions of lower
gas density where the radio plasma has been able to expand more.
On the western side of the radio mini-halo, the intensity gradient is 
steeper than that of the eastern side.
This may be due to 
gas compression by the S0 galaxy IC~5337 falling towards the cD 
galaxy IC~5338.
If this is the case, then IC~5337 should be very close to 
the central cooling core, rather than being part of the southwest subcluster.
This reinforces both our dynamical picture for the S0 galaxy and the 
interpretation of the mini-halo geometry by confinement.
The X-ray deficient regions in the NW and SE directions can also 
be seen on {\it XMM-Newton} residual map on a larger scale 
(dark regions of the left panel of Figure~\ref{fig:residual_xmm}).
The excess of intracluster gas density to the NE and SW might be due to
mergers along the axis of the local large scale structure, as discussed
in \S~\ref{sec:image_morph_outer},
or it may just represent the elongation of the equilibrium distribution
of the gas in this direction.
The lower density regions in the NE and SW directions can also be regions 
inflated by previous radio bubbles.
Although the symmetric radio bars may be due to the jet precession, it is
also possible that the denser gas to the NE and SW might explain the
shape of the two radio bars;
the radio bars might have been compressed radially by a collision with 
this gas, and become extended to the NW and SE by the smaller ICM density
gradients in those directions.

\section{Conclusions}
\label{sec:conclusion}

We have identified two extended X-ray emitting regions about $7\arcmin$
northeast 
and southwest from the center of Abell~2626, with the latter one 
probably associated 
with a subcluster identified to be falling into the main
cluster \citep{MGW96,MW97}.
We argue that the infalling velocity of the southwest subcluster is mainly 
along the line-of-sight.
Abell~2626 is known to be associated with the Perseus-Pegasus 
supercluster \citep{BB85a,BB85b,ZZS+93,EET+01}, 
and it is 
likely that the main cluster is accreting subclusters and groups from this 
large-scale structure 
(\S\S~\ref{sec:image_morph_outer}, \ref{sec:residual}, \& 
\ref{sec:spectrum:central}). 
From the hardness ratio map, both of the extended X-ray emitting regions 
appear to be harder than their surroundings, and each of them contains a 
softer point-like source.  The northeast point-like source may be 
associated with the galaxy 2MASX J23365372+2113322.

Based on the bow shape of the X-ray emission and the soft X-ray tail identified 
by the hardness ratio map, we argue that the S0 galaxy IC~5337
is falling towards the center of the main cluster from the west.
The steeper intensity gradient in the cluster radio mini-halo 
on the western side compared to the eastern side may also indicate that 
the radio plasma
is compressed by the S0 galaxy. This reinforces our interpretation that the 
S0 galaxy should be close to the cooling core region instead of 
located inside the southwest subcluster. 
On the other hand, if the S0 galaxy were associated with the subcluster in the
southwest direction, this motion would be hard to understand.
While there are two velocity measurements of IC~5337 consistent with the
southwest subcluster and three measurements consistent with the
main cluster, one possible explanation is that there may be two 
galaxies along the line-of-sight of IC~5337.

We have identified hard X-ray point sources due to AGNs associated with the 
cD galaxy IC~5338 and the S0 galaxy IC~5337, both of which are radio
sources.
The cD galaxy AGN is located on the southwest nucleus of the cD galaxy.
The S0 galaxy AGN shows evidence for excess internal absorption.

We have used a double beta model to fit the radial surface brightness profile
and generated residual images to understand 
the dynamics of the intracluster gas (\S~\ref{sec:residual}).
Most of the residual structures in the inner region occur on a 
scale similar to that of the radio sources in the center.  One major 
residual in the inner cluster which is not on the scale of the radio 
emission is the extended excess to the southwest.
It is located
between the cD galaxies and the southwest subcluster, and may be the
result of a merger.

The gas temperature in Abell~2626 increases from 
a central value of $\sim 2.5$~keV at $\sim 10$~kpc to a maximum of 
$\sim 3.5$~keV at around $230$~kpc,
and then decreases to
$\sim 1.5$~keV at $\sim 600$~kpc.
At a radius of $\sim 260$~kpc, there is a significant temperature jump 
seen in the fits to the {\it XMM-Newton} spectra.
The jump may be associated with a 
previous or an ongoing merger.
We used the X-ray spectra to determine the cooling rate within a 
projected 
radius of 72~kpc.
The amount of gas which is cooling down to very low temperatures is
less than a few $M_{\odot}$~yr$^{-1}$.
A much higher mass deposition rate of $\sim 74 \, M_{\odot}$~yr$^{-1}$ is 
obtained from the {\it XMM-Newton} spectra if the hot gas only cools 
by about a factor of three.

From the surface brightness and the temperature profiles, we have derived 
the density, pressure, and entropy profiles of Abell~2626.
At a radius of $\sim 70$~kpc, there is a slope discontinuity in the 
density and pressure profiles. 
The outer entropy profile has a logarithmic slope of $\sim 0.58$, which 
is lower than that of the universal scaling relation of $\sim 1.1$. A 
heating source is probably required to maintain the central entropy of 
$\sim 30$~keV~cm$^2$.
At a radius of $\sim 260$~kpc, there is a small bump in the entropy 
profile which may be associated with a merger.

Due to the nearly isothermal structure of Abell~2626 within the cooling 
radius, conduction is not important compared to cooling in the inner 
region.
The cooling radius is 
$\sim$70 kpc,  where the logarithmic slope of the surface 
brightness profile changes.   
At a radius of $\sim 260$~kpc, the conduction time scale is significantly 
shorter than the look-back time to a redshift of $z \sim 1$, which 
indicates that either the thermal structure (perhaps associated with a 
merger) happened recently or thermal conduction is highly suppressed.
We have also derived the total mass profile under the assumption of 
hydrostatic equilibrium, and compared the gas mass fraction with other 
relaxed clusters.
The global trend is typical for relaxed clusters.

We have investigated the cooling flow interaction with the 
central radio source.
The two elongated radio bars to the north and south of the center of the 
cD galaxy and the lack of obvious correlation between the two radio bars 
and any structures in the {\it Chandra} X-ray image, residual image, or 
hardness ratio map may indicate that the radio bars are thin tubes parallel to 
the plane of the sky.
Another possibility is that the radio plasma is mixed with the X-ray gas,
rather than displacing it.

The central radio structure may be due to jet precession.
The southwest cD nucleus is an AGN, and we suggest it has two 
precessing jets 
propagating towards the north and south, producing the two symmetric 
narrow radio bars.
The precession might be caused by the gravitational influence of the
second cD nucleus to the north.
The two jets may have uplifted cool gas from the central region, producing the 
X-ray excess ``tongues'' running from the center of the cD galaxy to the  
radio bars.
The northern tongue appears to be weaker, and it may be
disrupted by the northern nucleus of the cD galaxy.
Alternatively, the strength of the southern X-ray tongue may simply be due 
to the southern jet being stronger.
Jet precession may 
also be found in other cooling core clusters \citep{GFS06}.
It may help 
to solve the AGN feedback problem found in high resolution simulations 
\citep{VR06}.
Alternatively, the distortions of the radio source might be due to ICM
motion, particularly rotation, near the center of the cluster.

The radio mini-halo has a diamond shape, elongated in the northwest 
and southeast direction.
The diamond-shaped radio mini-halo seems to be compressed from the 
northeast and southwest direction, which is probably consistent with the 
merging scenario for Abell~2626.
With this compression geometry, the radio 
mini-halo might be leaking out in the northwest and southeast direction 
(Figure~\ref{fig:residual_chandra_halo}).  The western side of 
the mini-halo may also be compressed by the infalling S0 galaxy IC~5337.
Interestingly, the size of the mini-halo is roughly the same as the 
cooling radius, and is located at where the logarithmic slope of the 
surface brightness profile changes.  
The agreement between the size of the radio mini-halo and the cooling
radius
is consistent with the re-acceleration model for 
the origin of the mini-halo \citep{GBF+04}, although it might just be a 
coincidence since the definition of the cooling radius is somewhat arbitrary.
Independent of the origin of the radio mini-halo,
the change in the X-ray surface brightness slope at the outer edge of
the radio mini-halo may indicate that the radio plasma is
affecting the thermal plasma.

\acknowledgments
We would like to thank
H.~Andernach,
M.~N.~Chatzikos,
T.~E.~Clarke, 
M.~Gitti,
A.~M.~Juett,
J.~J.~Mohr,
S.~W.~Randall,
and
G.~R.~Sivakoff
for helpful discussions.
This work was supported by the National Aeronautics and
Space Administration, primarily through the {\it Chandra} award GO2-3160X
and through {\it XMM-Newton} award NAG5-13089,
but also through {\it Chandra} awards
GO4-5133X,
and
GO5-6126X,
and through
{\it XMM-Newton} awards
NNG04GO80G,
and
NNG06GD54G.
ELB was supported by a Clare Boothe Luce Professorship and 
by NASA through {\it Chandra} awards GO5-6137X and GO4-5148.
THR acknowledges support by the Deutsche Forschungsgemeinschaft through 
Emmy Noether Research Grant RE 1462 and by the German BMBF through the 
Verbundforschung under grant no.\ 50 OR 0601.

\begin{appendix}
%\appendix
\section{Density Deprojection}
\label{app1}

We basically followed \citet{KCC83} to deproject the gas distributions 
from surface brightness profile.
The procedure is briefly outlined here. 

Assuming spherical symmetric gas distribution, the observed X-ray surface 
brightness (in ergs cm$^{-2}$ sec$^{-1}$ Hz$^{-1}$ sr$^{-1}$) at an 
angle $\theta$ (corresponding to a projected physical 
radius $b$) from the center of the cluster is:
\begin{equation}
I_{\nu}^{\rm obs}(\theta) = \frac{1}{4\pi(1+z)^3} \int^{\infty}_{b^2} 
\frac{\epsilon_{\nu^\prime}(r) 
\, dr^2}{(r^2-b^2)^{1/2}} \, ,
\label{a1}
\end{equation}
where $\epsilon_{\nu}$ is the
X-ray emissivity of the gas integrated over all angles,
$ \nu$ is the observed frequency,
$ \nu^\prime = (1 + z) \nu$ is the emitted frequency,
$z$ is the redshift, and $r$ is the 
physical distance from the cluster center.

We divided the image into $n$ annuli bounded by the angular radii
$\theta_0 <  \theta_1 < \cdots < \theta_i < \theta_{i+1} < 
\cdots < \theta_n$ and centered on the peak of the cluster emission.  
In this paper, we take $\theta_0 = 0$.
Noting that $b = \theta \, d_A$, where $d_A$ is the angular diameter distance, 
the flux from the annulus from $\theta_i \to \theta_{i+1}$ is given by:
\begin{equation}
F_{\nu}^{\rm obs}(\theta_i \to \theta_{i+1}) = 
\int^{\theta_{i+1}}_{\theta_i} I_{\nu}^{\rm obs}(\theta) \, 2\pi \, \theta 
\, d\theta
= \frac{\pi}{d_A^2} \int^{b_{i+1}^2}_{b_i^2} I_{\nu}^{\rm obs}(b) \, db^2
\, .
\label{a2}
\end{equation}

Putting equation~(\ref{a1}) into (\ref{a2}), discretizing the integrals 
into summations, and assuming a constant emissivity 
$(\epsilon_{\nu^\prime , j})$ 
within each radius ($b_j \to b_{j+1}$),
the observed flux can be expressed as:
\begin{eqnarray}
F_{\nu}^{\rm obs}(\theta_i \to \theta_{i+1}) & = &
\frac{1}{3(1+z)^3 d_A^2} 
\left\{
\epsilon_{\nu^\prime , i} (b_{i+1}^2-b_{i}^2)^{3/2}
\right.
\nonumber
\\
& + & \sum_{j=i+1}^{n-1} \epsilon_{\nu^\prime , j} 
[ (b_{j+1}^2-b_{i}^2)^{3/2} - (b_{j}^2-b_{i}^2)^{3/2} 
\nonumber
\\
&   & 
\left.
-(b_{j+1}^2-b_{i+1}^2)^{3/2} + (b_{j}^2-b_{i+1}^2)^{3/2}
]
\right\}
\, ,
\label{a3}
\end{eqnarray}
or in matrix form:
\begin{equation}
F_{\nu}^{\rm obs}(\theta_i \to \theta_{i+1}) = \sum_{j=0}^{n-1} A_{ij} \, 
\epsilon_{\nu^\prime , j}
\, ,
\label{a4}
\end{equation}
where $A_{ij}$ contains the geometrical factors in equation~(\ref{a3}).  
The 
matrix $A_{ij}$ is triangular, and equation~(\ref{a4}) can easily be inverted
to give the emissivities at each radius.
Once the emissivity is known, it can then be converted to the electron 
density, $n_e$, by the relation:
\begin{equation}
\epsilon_\nu = n_e^2 \Lambda_\nu
\, ,
\end{equation}
where $\Lambda_\nu$ is the emissivity function which depends on the temperature
and the abundances.
Note that $\Lambda_\nu$ can be derived from the models used to fit the
X-ray spectra.

\section{Cooling Time Estimation}
\label{app2}
The cooling time of a region can be estimated from the normalization of
the MEKAL model fitted by XSPEC, together with the fitted temperature,
the bolometric luminosity estimated by XSPEC, and the deprojected
density.  
The normalization of the MEKAL model in XSPEC is given by:
\begin{equation}  
K = \frac{10^{-14}}{4\pi d_A^2(1+z)^2}\int{n_H n_e} dV
\, {\rm cm}^{-5} \, ,
\end{equation}  
where
$d_A$
and $z$ are the angular diameter distance and the redshift to
the source, respectively, $n_e$ and $n_H$ are the electron and hydrogen
densities, respectively, and $V$ is the volume of the region.  It is   
easy to show that the enthalpy $H$ of the region is:
\begin{eqnarray}
H & =      & \frac{5}{2}PV
\nonumber\\
  & \approx & \frac{2.5 \times 2.33 \times 4\pi d_A^2 (1+z)^2 10^{14} K
}{n_e}
kT
\, {\rm ergs} \, .
\end{eqnarray}
The instantaneous isobaric cooling time is then given by:
\begin{equation}
t_{\rm cool}^{*} = \frac{H}{L_{\rm bol}} \, ,
\end{equation}
where $L_{\rm bol}$ is the bolometric luminosity of the region which is
given by XSPEC.
The bolometric luminosity can be written as
$ L_{\rm bol} = \Lambda (T) n_e^2 V$, where the emissivity function
$\Lambda (T)$ depends on the abundances and is taken from the same
XSPEC model which best fits the spectrum.
Then, the integrated cooling time is 
\begin{equation}
t_{\rm cool} = t_{\rm cool}^{*} \, \frac{\Lambda (T)}{T^2} \,
\int_0^T \frac{ T^\prime \, d T^\prime }{\Lambda ( T^\prime )}
\approx \frac{t_{\rm cool}^{*}}{2} \, .
\end{equation}

\end{appendix}

\clearpage

\begin{deluxetable}{ccccccccc}
\tabletypesize{\scriptsize}
\tablewidth{0pt}
\tablecolumns{8}
\tablecaption{{\it XMM-Newton} X-ray Point Sources with NED Identifications
\label{table:source}
}
\tablehead{
\colhead{X-ray RA} &
\colhead{X-ray Dec} &
\colhead{NED object} &
\colhead{Type$^a$} &
\colhead{NED RA} &
\colhead{NED Dec} &
\colhead{Redshift} &
\colhead{Separation$^b$}
}
\startdata
23:36:04.8 & +21:06:11.2 &   2MASX J23360455+2106127 & G/IrS      &    23:36:04.6 & +21:06:13  & 0.066130 & 3\arcsec \\
23:36:09.6 & +21:20:25.1 &   MG3 J233615+2120        &RadioS      &    23:36:09.7 & +21:20:25  & \nodata & 1\arcsec \\
23:36:15.8 & +21:06:11.5 &   Abell 2626:[SPS89] 08   &VisS/RadioS &    23:36:16.2 & +21:06:21  & \nodata & 11\arcsec \\
23:36:24.5 & +21:09:03.6  &  IC 5337                  &G/IrS/RadioS&    23:36:25.1 & +21:09:02  & 0.054988 & 9\arcsec \\
23:36:30.5 & +21:08:47.0  &  IC 5338                  &G/IrS/RadioS&    23:36:30.6 & +21:08:50  & 0.054900 & 3\arcsec \\
23:36:53.8 & +21:13:32.9  &  2MASX J23365372+2113322  &G/IrS       &    23:36:53.7 & +21:13:32  & 0.056100 & 2\arcsec \\
23:36:54.2 & +21:15:31.7  &  2MASX J23365415+2115302  &G/IrS       &    23:36:54.2 & +21:15:29  & 0.054200 & 3\arcsec \\
23:36:57.1 & +21:14:05.6  &  2MASX J23365722+2114032  &G/IrS       &    23:36:57.2 & +21:14:03  & 0.038040 & 3\arcsec \\
23:37:32.4 & +21:09:07.6  &  2MASX J23373222+2109020  &G/IrS       &    23:37:32.2 & +21:09:02  & 0.184400 & 6\arcsec \\
\enddata
\tablenotetext{a}{
The source type as classified by NED:
G=Galaxies;
IrS=Infrared Source;
RadioS=Radio Source;
VisS=Visual Sources. 
}
\tablenotetext{b}{
Separation between the X-ray point source and source identified by NED.
}
\end{deluxetable}

\begin{deluxetable}{ccc}
\tablewidth{0pt}
\tablecolumns{3}
\tablecaption{
{\it Chandra} X-ray Point Sources
\label{table:source_chandra}
}
\tablehead{
\colhead{RA} &
\colhead{Dec} &
}
\startdata
23:36:15.82 & +21:08:55.4 & \\
23:36:21.31 & +21:13:29.7 & \\
23:36:21.41 & +21:11:35.4 & \\
23:36:24.61 & +21:08:47.7 & \\
23:36:24.81 & +21:10:16.7 & \\
23:36:25.04 & +21:09:03.0 & IC 5337 \\
23:36:25.60 & +21:13:07.7 & \\
23:36:30.52 & +21:08:47.0 & IC 5338 \\
23:36:33.30 & +21:08:36.7 & \\
23:36:36.98 & +21:10:51.8 & \\
23:36:37.46 & +21:06:21.0 & \\
23:36:38.43 & +21:08:26.1 & \\
23:36:45.60 & +21:07:49.3 & \\
23:36:46.84 & +21:08:07.8 & \\
\enddata
\end{deluxetable}

\clearpage

%\clearpage

\clearpage

\clearpage

%\clearpage

\end{document}